\documentclass{aa}  

\usepackage{graphicx}
\usepackage{txfonts}
\usepackage[colorlinks=true,citecolor=magenta]{hyperref}

\newcommand{\pcc}{\,pc\,cm$^{-3}$}

\begin{document}

   \title{Eighteen new fast radio bursts in the High Time Resolution Universe survey}   
   \titlerunning{Narrow-band FRBs in HTRU}


   \author{M. Trudu \inst{1} 
          \and
          A. Possenti \inst{1}  
          \and 
          M. Pilia \inst{1}  
          \and 
          M. Bailes \inst{2,3}  
          \and 
          E. F. Keane \inst{4} 
          \and      
          M. Kramer \inst{5,6}
          \and 
          V. Balakrishnan \inst{5} 
          \and 
          S. Bhandari \inst{7,8,9,10}
          \and 
          N. D. R. Bhat \inst{11}  
          \and
          M. Burgay \inst{1}
          \and 
          A. Cameron \inst{2,3}
          \and 
          D. J. Champion \inst{5} 
          \and   
          A. Jameson \inst{2,3} 
          \and 
          S. Johnston \inst{12}
          \and 
          M. J. Keith \inst{6}
          \and
          L. Levin \inst{6} 
          \and 
          C. Ng \inst{13}
          \and 
          R. Sengar \inst{14} 
          \and  
          C. Tiburzi \inst{1}      
          }

   \institute{
   INAF-Osservatorio Astronomico di Cagliari, via della Scienza 5, I-09047, Selargius (CA), Italy \and 
   ARC Center of Excellence for Gravitational Wave Discovery (OzGrav), Swinburne University of Technology, Mail H11, PO Box 218, VIC 3122 \and 
   Centre for Astrophysics and Supercomputing, Swinburne University of Technology, Hawthorn, Victoria 3122, Australia \and 
   School of Physics, Trinity College Dublin, College Green, Dublin 2, D02 PN40, Ireland \and  
   Max-Planck-Institut für Radioastronomie, Auf dem Hügel 69, D-53121 Bonn, Germany \and 
   Jodrell Bank Center for Astrophysics, University of Manchester, Alan Turing Building, Oxford Road, Manchester M13 9PL, United Kingdom \and 
   ASTRON, Netherlands Institute for Radio Astronomy, Oude Hoogeveensedĳk 4, 7991 PD Dwingeloo, The Netherlands \and 
   Joint institute for VLBI ERIC, Oude Hoogeveensedĳk 4, 7991 PD Dwingeloo, The Netherlands \and 
   Anton Pannekoek Institute for Astronomy, University of Amsterdam, Science Park 904, 1098 XH, Amsterdam, The Netherlands \and 
   CSIRO Astronomy \& Space Science, Australia Telescope National Facility, P.O. Box 76, Epping, NSW 1710, Australia \and 
   International Centre for Radio Astronomy Research, Curtin University, Perth, WA, Australia \and 
   Australia Telescope National Facility, CSIRO Space and Astronomy, PO Box 76, Epping NSW 1710, Australia \and 
   Laboratoire de Physique et Chimie de l’Environnement et de l’Espace - Universit\'e d’Orl\'eans/CNRS, 45071, Orl\'eans Cedex 02, France \and 
   Center for Gravitation, Cosmology, and Astrophysics, Department of Physics, University of Wisconsin-Milwaukee, PO Box 413, Milwaukee, WI 53201, USA  
             }

   \date{Received September 15, 1996; accepted March 16, 1997}


 
  \abstract
   {Current observational evidence reveals that fast radio bursts (FRBs) exhibit bandwidths ranging from a few dozen MHz to several GHz. Traditional FRB searches primarily employ matched filter methods on time series collapsed across the entire observational bandwidth. However, with modern ultra-wideband receivers featuring GHz-scale observational bandwidths, this approach may overlook a significant number of events. }   
   {We investigate the efficacy of sub-banded searches for FRBs, a technique seeking bursts within limited portions of the bandwidth. These searches aim to enhance the significance of FRB detections by mitigating the impact of noise outside the targeted frequency range, thereby improving signal-to-noise ratios.}
   {We conducted a series of Monte Carlo simulations, for the $400$-MHz bandwidth Parkes 21-cm multi-beam (PMB) receiver system and the Parkes Ultra-Wideband Low (UWL) receiver, simulating bursts down to frequency widths of about $100$\,MHz. Additionally, we performed a complete reprocessing of the high-latitude segment of the High Time Resolution Universe South survey (HTRU-S) of the Parkes-Murriyang telescope using sub-banded search techniques.}   
   {Simulations reveal that a sub-banded search can enhance the burst search efficiency by $67_{-42}^{+133}$\,\% for the PMB system and $1433_{-126}^{+143}$\,\% for the UWL receiver. Furthermore, the reprocessing of HTRU led to the confident detection of eighteen new bursts, nearly tripling the count of FRBs found in this survey. }
   {These results underscore the importance of employing sub-banded search methodologies to effectively address the often modest spectral occupancy of these signals.}

   \keywords{methods: data analysis --
                fast radio bursts 
               }

   \maketitle
%



\section{Introduction}
\label{sec:intro}
Fast radio bursts (FRBs) are Jy-intensity radio flashes with milliseconds durations, primarily exhibiting dispersion measures (DMs) that greatly exceed the contribution from the Galaxy \citep{htrufrb1}, being hence (mostly) extragalactic sources \citep[for recent reviews, see, e.g.,][]{bailes_2022_review, petroff_2022_frbreview}. After the initial discovery of the first FRB event by \cite{lorimer_2007_firstfrb}, the first population of bursts was reported in 2013 by \cite{htrufrb1}. This discovery was made by analysing a subset of data from the high-latitude portion of the High Time Resolution Universe \citep[HTRU,][]{htru1} survey conducted using the 64-m single-dish Parkes-Murriyang telescope in Australia. Subsequently, the field experienced rapid growth. Notably, the identification of repeating sources \citep[e.g.,][]{spitler16, chime_2019_8newfrb} facilitated dedicated large-scale campaigns, enabling precise localisation and host galaxy identification \citep[][]{marcote17,tendulkar17,marcote20,kirsten_2022_20e_loc}, the discovery of periodic activity in two repeaters \citep[][]{chime_2020_periodr3,rajwade20}, and the detection of emission with nanosecond duration \citep[][]{nimmo_2021_r3nanoshots,nimmo_2022_20Enanoshots,walid_2021_20Enanoshots,snelders_2023_r1nanoshots}.

The blind search for these transient events is usually done by searching for excesses in the time series of the recorded signal via matched filtering \citep{cordes_2003_sp,qiu_2023_frbsearch}. After extracting  the Stokes I spectrogram from the original complex voltage of the signal, the essential steps of a FRB search pipeline are the following: (i) initially cleaning the data to remove radio frequency interference (RFI) signals that can disrupt the burst search if not properly eliminated; (ii) subsequently adjusting the Stokes I matrix for several trial DMs by shifting each channel row according to the corresponding DM-induced delay; (iii) assuming a flat spectral index for the FRB emission and averaging the DM-corrected matrices in frequency; (iv) convolving the time series with top-hat functions using various trial boxcar widths, while retaining candidates above a given signal-to-noise ratio (S/N) threshold; (v) clustering temporally coincident candidates with matching DM/boxcar widths into a single event; (vi) vetting the grouped events with either an Artificial Intelligence (AI) classifier \citep[see, e.g.,][]{connor_2018_ml, fetch} or by human inspection. 

In contrast to pulsars, which display broad fractional bandwidths \citep{jankowski_2018_spr_spec}, in some extreme cases extending to $100$\,GHz \citep{torne_psr_millimiter}, the spectral occupancy of FRBs appears in many case to be narrower. This effect seems particularly pronounced for the repeater class, rather than for the one-offs, suggesting the possibility of a morphological dichotomy between the two classes of FRBs \citep{pleunis_2021_frbmoprh}.

Furthermore, the repeaters' bandwidths appear to be frequency-dependent. For instance, FRB\,20121102A exhibits bandwidths of a few hundred MHz the $1.4$\,GHz \citep{hewitt_2021_r1storm} and broadens to the order of gigahertz at $6$\,GHz \citep{gajjar_2018_r18ghz}. A similar trend is observed for FRB\,20180916B, where bursts at $0.6$\,GHz show bandwidths of hundreds of MHz \citep{sand22_multiband}, reaching GHz spans at $4.5$\,GHz \citep{bethapudi_2022_r3_4ghz}. 

To conclusively determine if this dichotomy in bandwidths between repeaters and one-off events exists, observations with receivers possessing extremely large observational bandwidths could provide a definitive answer. If these results hold for larger samples of bursts, they could also help answer the question of whether repeating FRBs and one-off FRBs constitute separate classes of events, as this distinction might not exist \citep{james_2023_frballrepeaters}. 

Due to the narrower spectral occupancy of FRBs compared to the full observational bandwidth, searching for signal excesses in the frequency-averaged time series of the entire band might be expected to introduce excessive noise. This situation could potentially cause the burst signal to fall below the S/N threshold. This issue is particularly pronounced with modern receivers like the Parkes Ultra-Wideband Low \citep[UWL,][]{uwl}, boasting an observational bandwidth of approximately $3.3$\,GHz. A practical approach aimed to not miss events which occupy a significantly smaller portion of the full observational bandwidth involves conducting a sub-banded search. This method entails searching for bursts in a manner akin to the previous discussion, but focusing on smaller portions of the data matrix in terms of frequency. This process is then iteratively applied across all sub-bands. For instance, this technique enabled \cite{kumarsub} to identify a very narrow burst with a bandwidth of about $65$\,MHz from the repeater FRB\,20180711A within the UWL data.

This work aims to provide a comprehensive analysis of sub-banded burst searches, focusing particularly on demonstrating the potential gain of the number of the detections. To illustrate this concept, we conducted a Monte Carlo simulation and we also performed a reprocessing of the high-latitude portion of the HTRU survey, resulting in the discovery of 18 new FRBs. The structure of this paper is as follows: Sect.\,\ref{sec:subbandsearches} outlines the framework for sub-banded searches and a description of our simulations; Sect.\,\ref{sec:htrusurvey} provides a concise overview of the HTRU survey and details the software pipeline we developed for processing; Sect.\,\ref{sec:htruresults} presents the outcomes of our reprocessing; and finally, Sect.\,\ref{sec:conclusions} provides a summary and concluding remarks.


\section{Sub-banded search algorithm}
\label{sec:subbandsearches}

\subsection{Design} 
\label{subsec:algodesign}

\begin{figure}
    \centering
    \includegraphics[width = \linewidth]{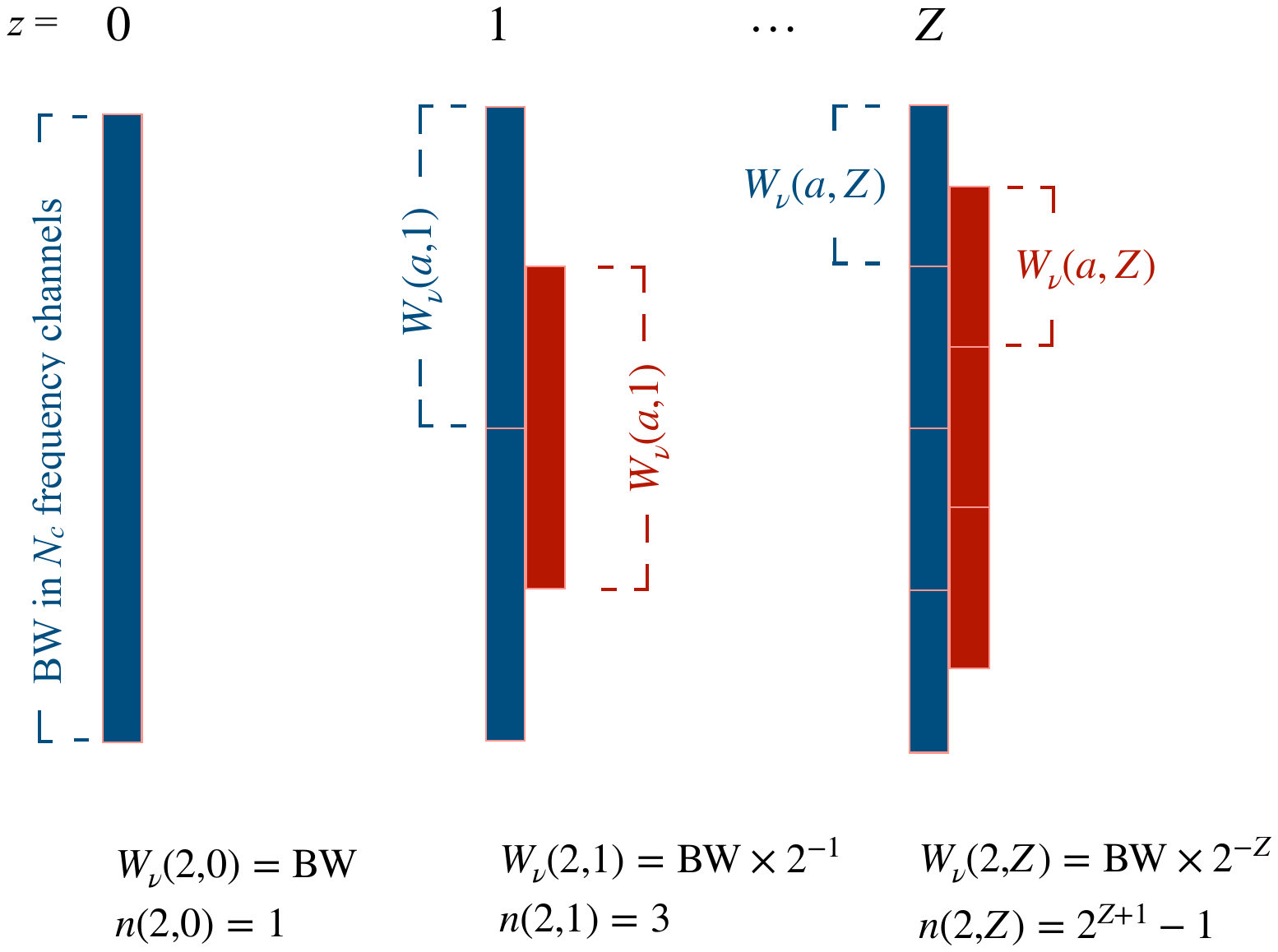}
    \caption[]{Strategy for the FRB sub-banded search (see Sect.\,\ref{subsec:algodesign}). The full band BW is iteratively divided up to sub-bands $W_{\nu}(a,Z)$, given a certain partition factor and partition exponent (represented here as blue rectangles). Overlapping sub-bands (red rectangles) are considered in order to search for events which occupy adjacent sub-bands. }
    \label{fig:algodesign}
\end{figure}

The design of a sub-banded search algorithm can be done in various ways, each tailored to specific scientific results one wants to achieve. One possible approach involves segmenting the entire observational bandwidth into distinct sub-bands of varying sizes. If the burst is fully contained in a given sub-band $W_{\nu}$, assuming a Gaussian shaped spectrum with FWHM $W_{\nu}$, it can be shown that (see Appendix\,\ref{subsec:htrusnr}):
\begin{equation}
    \label{eq:snrgainfinal} 
    {\rm S/N}^* \simeq \sqrt{\frac{{\rm BW}}{W_{\nu}}} {\rm S/N} \ ,
\end{equation}
where S/N$^*$, S/N are the signal-to-noise ratios in the given sub-band and in the full band BW respectively.
In this context, we present a procedural framework for sub-banded searches, visually outlined in Fig.\,\ref{fig:algodesign}. We consider a channelised spectrogram with $N_c$ frequency channels, within an observational band BW. We evenly divide the total observational bandwidth into $a^z$ sub-bands, where $z = 0,1,...,Z$, with $a$ the partition factor and  $Z$ the partition exponent.  
Consequently, for each $z$, each sub-band possesses a bandwidth:
\begin{equation}
    \label{eq:bwsub} 
    W_{\nu} (a,z) = {\rm BW} \times a^{-z} \ .
\end{equation}
As depicted in Fig.\,\ref{fig:algodesign}, we also consider for each $z$ adjacent sub-bands equally large $W_{\nu} (a,z)$. This is done in order to mitigate the possibility of missing events that might manifest between adjoining sub-bands. The number of sub-bands $n(a,z)$ produced in a given exponent $z$ is simply:
\begin{equation}
    \label{eq:nsub} 
    n(a,Z) = 2 a^{z} - 1 \ . 
\end{equation}

The overall count of sub-bands $N(a,Z)$ to be processed can be calculated as:
\begin{equation}
    \label{eq:nsubtot} 
    N(a,Z) = \sum_{z=0}^{Z} n(a,z) = \frac{2 a^{Z+1} - a (Z+1) + Z - 1}{a-1} \ . 
\end{equation}

To exemplify, let us consider $a = 2$ and $Z=2$. This configuration encompasses the entire observational band BW, three bands each constituting half of BW, and seven bands each occupying a quarter of BW. Consequently, a total of 11 bands necessitate processing. 

The choice of the parameters $a$ and $Z$ requires a compromise between the desired sub-band width $W_{\nu} (a,Z)$ one wants to search for bursts and the manageable count of total searches $N(a,Z)$, as it exponentially increases for large values of $Z$.


\subsection{Detection gain} 
\label{subsec:gain} 

\begin{figure}
    \centering
    \includegraphics[width = \linewidth]{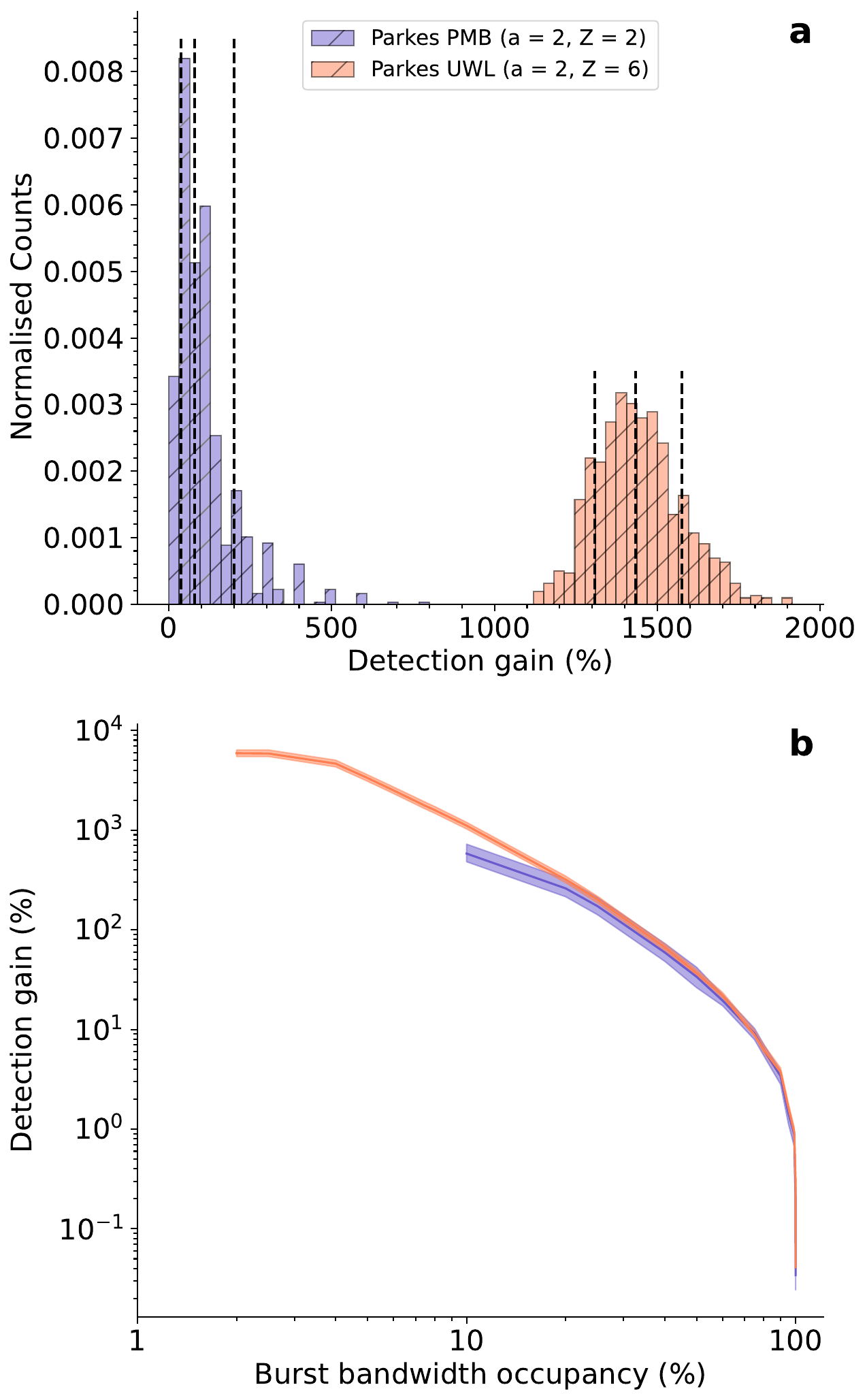}
    \caption[]{Results of the Monte Carlo experiments (see \ref{subsec:gain} for more details). The top panel (a) displays the distribution of the detection gain obtained by performing a sub-banded search over a full-band search (Experiment 1), both for the PMB (violet, search parameters $a = 2$ and $Z = 2$) and UWL (coral, search parameters $a = 2$, $Z = 6$) receivers. The dashed vertical lines represent the distribution percentiles at 16, 50, and 84\%. The bottom panel (b) illustrates the sub-banded search detection gain as a function of the burst bandwidth occupancy (Experiment 2). The colour code and the sub-banded search parameters are the same as the top panel.}
    \label{fig:gain}
\end{figure}

To assess the increased detection capability of the sub-banded search discussed in Sect.\,\ref{subsec:algodesign}, we conducted two Monte Carlo experiments. These simulations focused on two receivers: the Parkes 21-cm multi-beam receiver system \citep[PMB,][]{parkes_multibeam}, recording data at the central frequency of $1382$\,MHz with an observational bandwidth of $400$\,MHz, and the UWL, centered at $2368$\,MHz within a $3328$\,MHz band. In order to have comparable results with the real data application we will discuss in Sect.\,\ref{subsec:htrupipeline}, we consider for the PMB an effective bandwidth of $340$\,MHz. 

In both simulations, an FRB event was generated as a Gaussian function $G(\nu; \{ \nu_0, \Delta {\nu}, F_{\nu} \})$. This function is characterised by three parameters: $\nu_0$, the central emission frequency corresponding to the Gaussian centroid; $ \Delta \nu $, the Gaussian FWHM representing the burst frequency width; and $ F_\nu $, the burst fluence, calculated as $ \int G(\nu )  d \nu $. 

In the sub-banded search, the fluence is evaluated within limited frequencies  $\nu_1, \nu_2$, as described in Sect.\,\ref{subsec:algodesign}. If the fluence $\int_{\nu_1}^{\nu_2} G(\nu) d \nu$  in one of the sub-bands exceeds the telescope radiometer fluence sensitivity (with a S/N threshold of 10), improved by a factor of ${\rm BW} / (\nu_2 - \nu_1)  $ (see Eqs.\,\ref{eq:radiometer1},\ref{eq:radiometer2}), the event is labelled as detected.

For both experiments, some parameters are drawn randomly, while others are kept fixed. Below, we describe the simulation parameters.

    Experiment 1: We produced $10^4$ bursts, whose parameters $\{ \nu_0, \Delta {\nu}, F_{\nu} \}$ are randomly generated. Fluences are drawn from a power-law distribution between the receiver's sensitivity fluence of $0.6$\,Jy\,ms and $10^3$\,Jy\,ms, assuming a slope of $-3/2$. Considering current evidence for FRB bandwidths, we generated widths uniformly $\mathcal{U}$ distributed in the range $70-1000$\,MHz for both receivers. Frequency centroids are drawn from uniform distributions $\mathcal{U} (400,1900)$\,MHz  and $\mathcal{U} (400,4400)$\,MHz  for the PMB and UWL, respectively. The ranges are slightly larger than the receiver's band in order to consider outside-band events whose spectral widths are, however, large enough to significantly occupy the receiver's band. For each event, we performed a full-band search and a sub-banded search with parameters $a = 2, Z = 2$ for the PMB and $a = 2, Z = 6$ for the UWL. After counting the number of detected bursts for both searches and receivers, we repeated the experiment for $10^3$ trials. This experiment aims to provide a prediction of the gain in detections when performing a sub-banded search compared to a full-band search, given a certain sub-banded search setup, for a given dataset, assuming no prior knowledge of the bandwidth occupancy of the putative bursts in the data.
 
    Experiment 2: In this case, we keep fixed the percentage bandwidth occupancy of the bursts (henceforth the spectral widths), considering the following bandwidth occupancies: $[10,20,25,40,50,60,75,80,90,95 ,99,99.9,99.99]$\,\% for the PMB and $[1,2,2.5,4,5,6,7.5,8,10,20,25,40,50,60,75,\\80,90,95,99,$ $99.9,99.99]$\,\%  for the UWL. For each fixed occupancy, we generated $10^4$ bursts, with fluences distributed as previously described and with frequency centroid uniformly distributed within the receiver bandwidth, ensuring that each burst is fully contained within the band. We again performed a full-band search and a sub-banded search with the same parameter setup, counted the detected events, and repeated for $10^3$ trials for each spectral occupancy. With this experiment, we wanted to evaluate the detection gain as a function of the burst percentage occupancy.


Figure\,\ref{fig:gain} illustrates the results of the two Monte Carlo experiments. In the top panel, the detection gain distributions from Experiment 1 for both PMB and UWL are displayed. The UWL gain distribution is approximately Gaussian, centred around $\sim 1400$\,\%, while the PMB gain distribution is positively skewed, with a median gain of $67$\,\%. This skewness in the PMB gain distribution is expected, given our priors for the width distribution. A substantial portion of the generated events occupies the entire observational bandwidth of the PMB, shifting the gain toward lower values. In contrast, for the UWL, this scenario never occurs for construction, as the widest bursts would only occupy about $30$\,\% of the UWL bandwidth.

From these results, we can establish an overall detection gain for both receivers, considering the sub-banded search setup used. Assuming the $16-84$ percentile range of our distribution as a $1 \sigma$ uncertainty, we obtain an overall detection gain of $\mathcal{G}_{\rm PMB} = 67_{-42}^{+133}$\,\% for the PMB and $\mathcal{G}_{\rm UWL} = 1433_{-126}^{+143}$\,\%.

The bottom panel of Fig.\,\ref{fig:gain} depicts the detection gain as a function of burst occupancy. For very low occupancies, such as the case of $2$\,\% for the UWL \citep[like the burst detected by][]{kumarsub}, the detection gain becomes remarkably high, reaching $\sim 6000$\,\%. This implies that, given a certain UWL dataset, if a full-band search detects a single event, a sub-banded search with $a = 2$ and $Z = 6$ could potentially allow the detection of 60 more events. Conversely, as we approach occupancies greater than $90$\,\%, the detection gain for both receivers dramatically decreases to zero.


\section{Reprocessing of the HTRU South survey}
\label{sec:htrusurvey}
\subsection{Observations}
\label{subsec:surveyspec}  

As a real-data application we reprocessed the HTRU survey by using the sub-banded search algorithm. The HTRU survey is an all-sky survey which was designed for discovering pulsars and fast transients. It consists of two parts: one carried out with the Parkes-Murriyang telescope which covered the Southern hemisphere of the Sky (HTRU-S) and the other made by the 100-m single dish Effelsberg telescope in Germany to cover the Northern Sky (HTRU-N). In this work we focus only on HTRU South and we refer the readers to \citet{htrun1} for a complete overview of the observational setup and strategy for HTRU North. 

The HTRU South survey is divided into three surveys, which cover three different regions of the Sky: the low-latitude (LowLAT) survey, which covers the sky area of Galactic longitude $-80^{\circ} < l < +30^{\circ}$ at Galactic latitudes $-3.5^{\circ} < b < +3.5^{\circ}$; the mid-latitude (MedLAT) survey, which comprises the region of $-120^{\circ} < l < +30^{\circ}$ and 
$-15^{\circ} < b < +15^{\circ}$ and lastly the high-latitude (HiLAT) survey, which covers the entire region of the Southern Sky with declination $\delta <10^{\circ}$. In this work we focused on only HiLAT as being the one designed for extra-galactic radio transients. 

The data were recorded with the PMB receiver. The PMB consists of 13 feeds centred around the prime focus of the Parkes antenna organised as two concentric sets of hexagons. The HTRU HiLat data are recorded as 2 bits total-intensity search-mode {\sc sigproc} filterbanks \citep{sigproc}. Each filterbank file possesses 1024 frequency channels of bandwidth $d \nu = 390.626$\,kHz and sampled at $dt = 64$\,$\mu$s each. Each observation lasted $270$\,s. 

\subsection{Data analysis}
\label{subsec:htrupipeline}
\begin{figure}
    \centering
    \includegraphics[width = \linewidth]{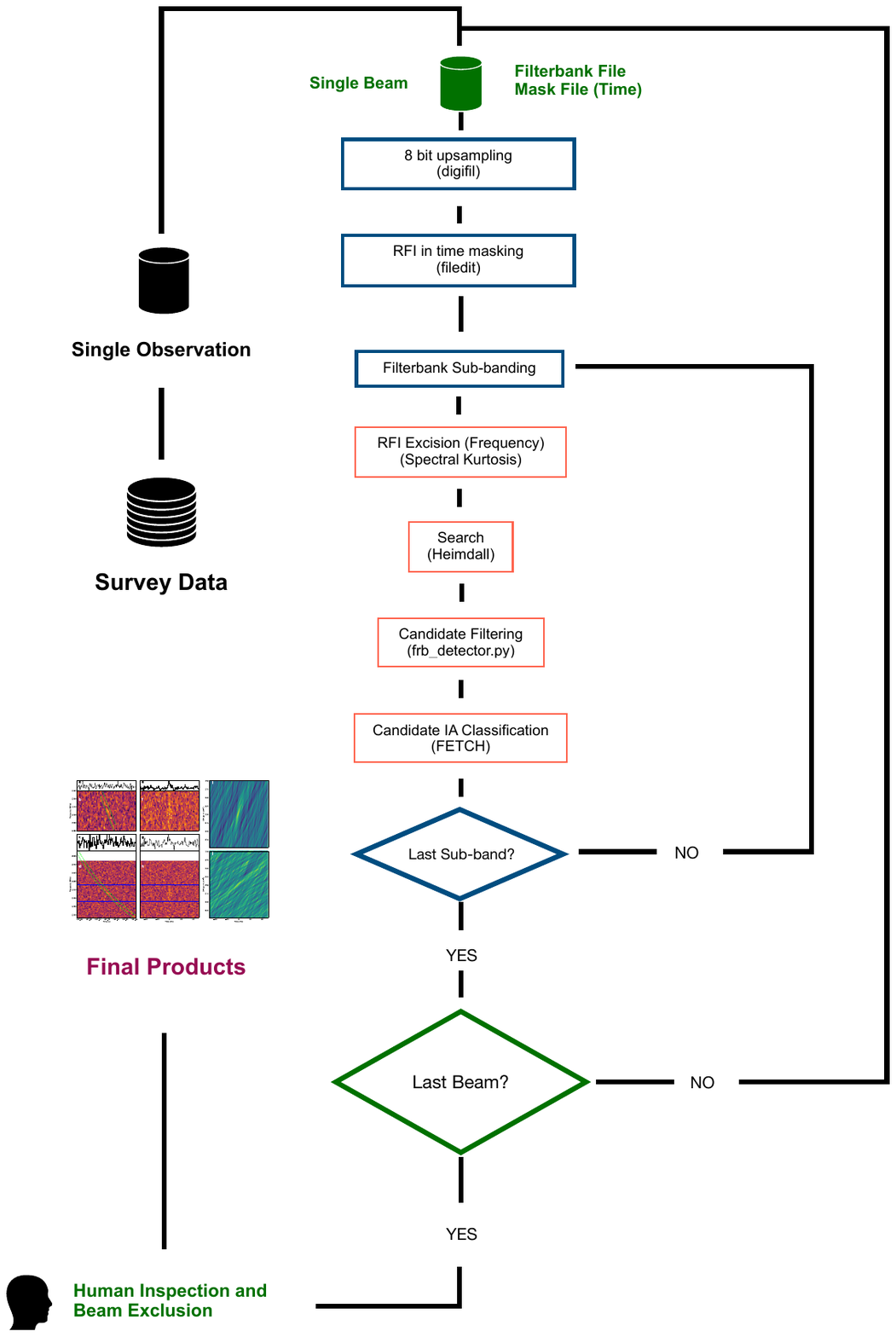}
    \caption[Flow chart of the sub-band search pipeline]{Flow chart of the sub-band search pipeline deployed to process the HTRU HiLAT.}
    \label{fig:htruflowchart} 
\end{figure}

To process the HiLAT segment of the survey, we developed a dedicated pipeline, outlined in Fig.\,\ref{fig:htruflowchart}. Data processing was executed using the OzSTAR supercomputer\footnote{\url{https://supercomputing.swin.edu.au/ozstar/}}, hosted at Swinburne University in Australia. Each observation comprised 13 individual {\sc sigproc} filterbank files, one for each beam of the PMB receiver. The processing sequence for each filterbank file involved the following steps:

\begin{enumerate}
    \item File upsampling. As a first step, the filterbank file is upsampled from a 2\,bits file to a 8\,bits file by using the software routine {\sf digifil} from the {\sc dspsr} software package \citep{dspsr}. This step was necessary in order to make the data readable by the AI classifier at later stages.
    \item  RFI excision in time. Accompanying each filterbank file was an RFI mask detailing the worst-affected time bins \citep[see ][Sect.\,4.1.1]{htru1}. Corrupted bins were replaced with random Gaussian numbers drawn from the distribution of uncorrupted time bins. This process was executed via the {\sf filedit} routine from {\sc sigproc}.
    \item File sub-banding and processing. Each file is sub-banded according to the procedure that was outlined in Sect.\,\ref{subsec:algodesign} and we considered $a = 2$ as partition factor and $Z = 2$ as partition exponent, which yielded the following configuration of  sub-bands: $1 \times 400$\,MHz, $3 \times 200$\,MHz and $7 \times 100$\,MHz. This was done in order to compromise between search sensitivity and computation time. For each sub-band we then processed the data according to the following steps:
    
        (i) For each sub-banded file we search for the noisiest channels present in the data. We adopted a Spectral Kurtosis algorithm \citep[provided by the python package {\sc your},][]{your}. As previously mentioned in Sect.\,\ref{subsec:gain}, about $60$\,MHz ($\sim 154$ channels) of the top band is known to be always affected by RFI due to the presence of satellite telemetry. These channels have been always flagged as zero. 
        
        (ii) FRB candidates have been searched via the software package {\sc heimdall} \citep{bbb+12}. As {\sc heimdall} performs its own RFI flagging we parsed the previously computed corrupted channels via the option \texttt{-zap\textunderscore chans} to ensure a further RFI excision. The data were searched in DM from $0$ to $5000$\,\pcc, over box-car widths from $0.128$\,ms (2 bins) to $262.144$\,ms (4096 bins).
        
        (iii) Before using an AI classification we made a pre-filtering step to significantly reduce the sheer amount of candidates produced by {\sc heimdall} by matching the following criteria: 
        \begin{equation} 
        \label{eq:htrusifting}
        \begin{split}
        {\rm S/N}    & \geq 7  \ ; \\
        \textrm{DM}     & \geq 10 \ \textrm{\pcc} \ ; \\ 
        N_{m}  & \geq 3 \ ; \\
        N_{1 {\rm s}} & \leq 2  \ .  
        \end{split}
        \end{equation}
        Where $N_m$ is the minimum number of distinct boxcars/DM trials clustered into a single candidate by {\sc heimdall} and $N_{1 \rm s}$ is the maximum number of candidates permitted in a 1 second-long window. The last two criteria were used to mitigate events most likely caused by noise fluctuations and RFI storms, respectively. However, the latter could have filtered out second-long events like those reported by CHIME \citep{michilli_2022_subsecfrb}, a trade-off we accepted between missing some events and reducing the number of probable false positives passed to the classifier. This selection process was implemented using the software \texttt{frb\textunderscore detector.py} \citep{bbb+12}.

        (iv) Subsequently, candidates satisfying the criteria detailed in Eq.\,\ref{eq:htrusifting} were subjected to scrutiny by {\sc fetch} \citep{fetch}. {\sc fetch} is an AI classifier which offers eleven convolutional neural network architectures (referred to as models {\it a} to {\it k}), each with a distinct configuration of layers \citep[for details, see][]{fetch}. Model {\it a} was exclusively employed, as it was reported by the authors as the most effective overall.

    \item Human evaluation. Lastly the candidates positively classified by {\sf FETCH} are humanly evaluated.  We describe the criteria we used. As a conservative approach all the candidates which showed bright clustered pixels, in the dynamic spectrum (additional to the clustered ones of the putative FRB candidate) were discarded. This is done conservatively to avoid possible RFI signals that are temporally too close to the single pulse candidates. Candidates with a DM compatible with the Galactic DM predicted by using the NE2001 \citep{ne2001} model and YMW16 \citep{ymw17} for the beam sky direction were also rejected, their vetting (e.g. positively classify them as new pulsars or RRATs) will be part of a future work which will comprise the reprocessing of LoLAT and MedLAT. Candidates which appearead in multiple but not adjacent beams, according to the geometry of the PMB were not considered. 

\end{enumerate}

\section{Results and discussion}
\label{sec:htruresults}

\subsection{Bursts re-detected}
\begin{figure}
    \centering
    \includegraphics[width = \linewidth]{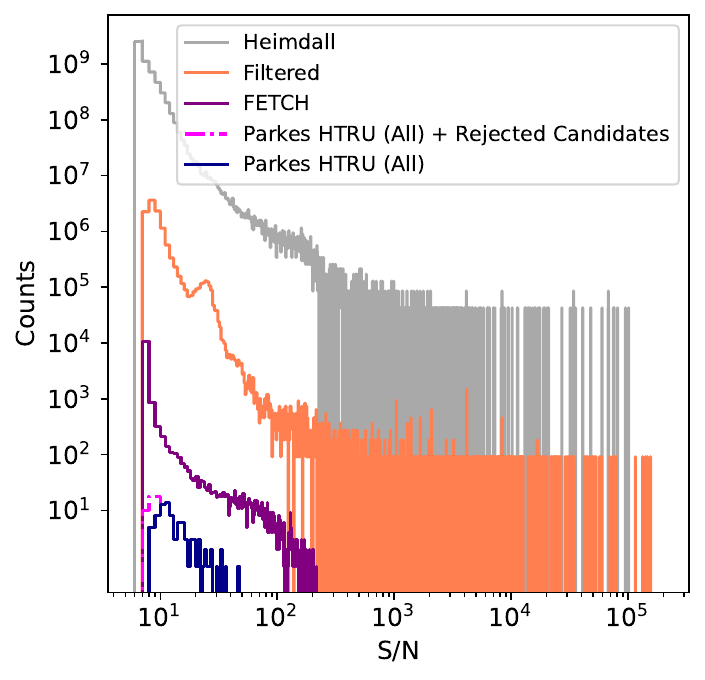}
    \caption[]{S/N distributions (binned with an S/N bin size of 1) of the candidates: in grey, those initially detected by {\sc heimdall}; in coral, the {\sc heimdall} candidates filtered according to the criteria listed in Eq.\,\ref{eq:htrusifting}; in purple, the filtered candidates positively classified by {\sc fetch}; in magenta, the sample of our 51 detections along with the 10 previously discovered FRBs in this survey \citep[][including detections in multiple sub-bands]{htrufrb1,htrufrb2,htrufrb3}; and in blue, the same distribution as the previous one, but with the sample of our detections limited to events with $\rm S/N \geq 10$ (see Sect.\,\ref{subsec:plausibility} for more details).
}

    \label{fig:snrdistr}
\end{figure}

Following the complete processing, {\sc heimdall} identified approximately $6 \times 10^9$ candidates (above a S/N$\geq$ 6), of which approximately $10^7$ survived the pipeline filtering steps as discussed in Sect.\,\ref{subsec:htrupipeline}. Only $\sim 10^4$ of these were positively classified by {\sc fetch} (See Fig.\,\ref{fig:snrdistr}). 



\begin{figure}
    \centering
    \includegraphics[width = \linewidth]{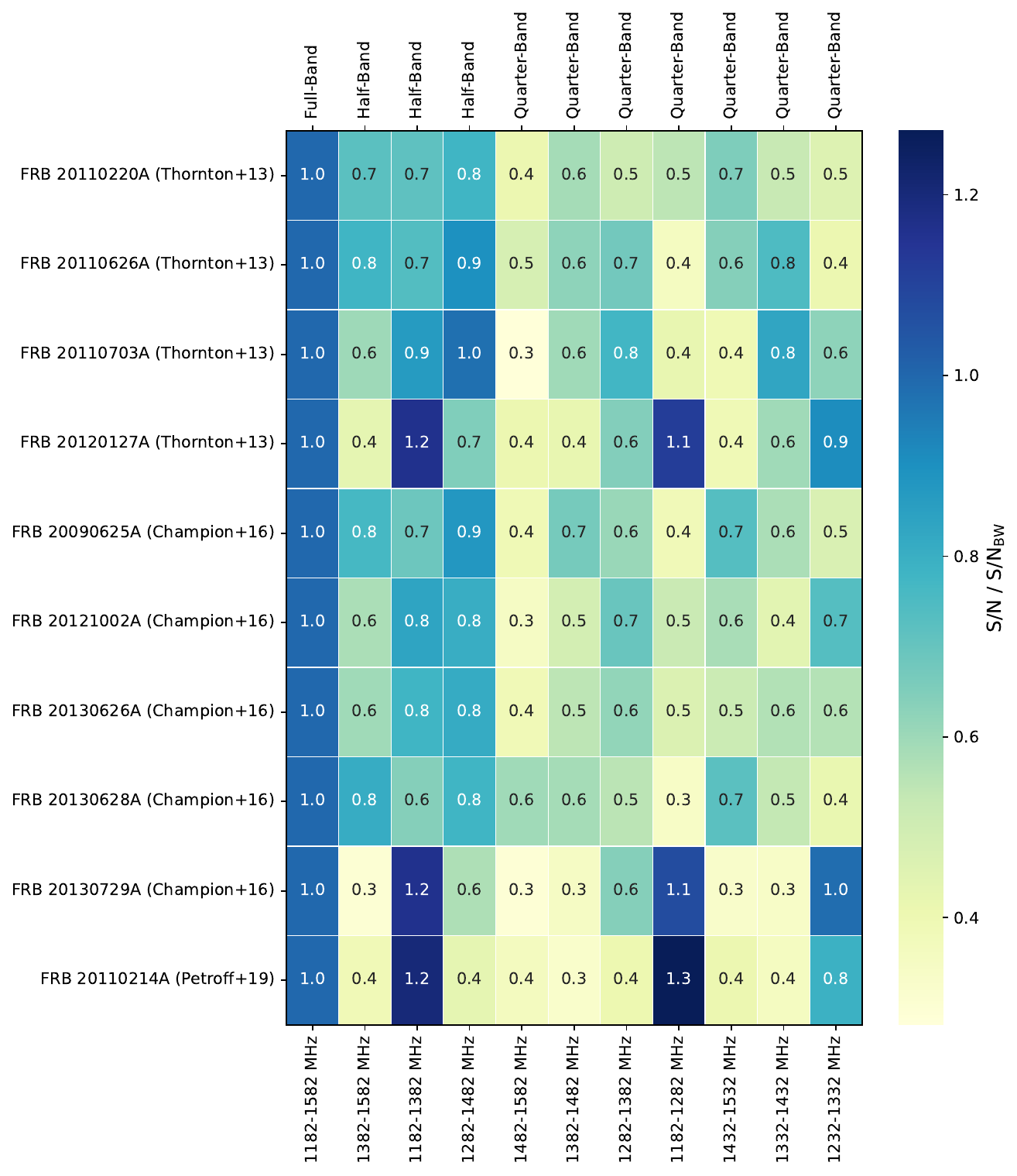}
    \caption[]{Recomputed relative S/N (normalised for the full-band S/N) for each sub-band processed in the search and for each of the previously detected HTRU FRBs. For each burst, the relative sub-band has been considered, and the S/N computed using Eq.,\ref{eq:snrgain4}.}
    \label{fig:snrmap}
\end{figure}

We re-detected the 10 previously discovered FRBs in the HiLAT region \citep{htrufrb1, htrufrb2, htrufrb3}, both in the full-band and across several sub-bands. Notably, the S/N values of these sub-banded bursts, such as those from FRB\,20110220A, provide significant insights. The burst in the full-band exhibits an S/N of approximately 49. Interestingly, this burst has an S/N of $\sim 70$,\% of the full-band S/N if we consider half sub-bands (see Fig.,\ref{fig:snrmap}) and about $50$,\% of the full-band S/N in the quarter sub-bands. We observe hence a S/N loss rather than a gain. A similar scenario applies for FRB\,20130626A. In contrast, we observe that FRB\,20110127A, 20130729A, and 20110214A exhibit a higher S/N in the lower half-band, as the majority of their signal is concentrated in this frequency range \citep[this was already noted for FRB\,20110214A, see][]{htrufrb3}. With the exception of these three bursts, the spectra of all published FRBs span the entire observational bandwidth. In such instances, it can be shown (see Appendix\,\ref{subsec:htrusnr}) that, if the burst fully occupies the observational bandwidth, Eq.\,\ref{eq:snrgainfinal} assumes the form:
\begin{equation}
\label{eq:snrloss}
{\rm S/N}^* \simeq \sqrt{\frac{W_{\nu}}{{\rm BW}}} {\rm S/N} \ .
\end{equation}
This conclusion is consistent with the aforementioned bursts from \cite{htrufrb1} and \cite{htrufrb2}, and it reasonably applies to all 7 full-band HiLAT bursts, albeit with certain discrepancies that could arise mainly due to the presence of RFI or the limitations of the burst model. Another factor could be the assumption that, as initially discussed in Sect.,\ref{sec:intro}, we assumed a flat spectral index for the FRB emission, i.e., when collapsing the Stokes I matrix to get the time series to be searched for, we do not weight each channel of central frequency $\nu$ for a factor of the kind $(\nu/\nu_0)^{\beta}$, where $\nu_0$ is a reference frequency and $\beta $ the spectral index. To test this, we considered FRB\,20110220A, which is the strongest burst of the HTRU sample, and recomputed the S/N of the burst by considering trial spectral index values in the range of (-10,10). We notice that the S/N peaks at $\beta \simeq 2$ however, it results in an S/N improvement of less than $1$\,\%.

\subsection{New detections}
\label{subsec:plausibility}


Among the $10^4$ candidates that successfully passed the {\sf FETCH} selection and excluding the previous 10 discovered FRBs, only 51 fulfilled the criteria outlined in the human vetting section of the pipeline. Our identified candidates exhibit sub-band S/N values ranging from 7 (the minimum allowable value according to our filtering criteria) to 12.

We first discuss the probability that these bursts are simply due to random excesses in the frequency averaged time series. \cite{cordes_2003_sp} showed that, for a time series of $N_s$ samples, the average number of false detections $n_{\rm false}(> {\rm S/N})$ by chance due only to noise, for a single DM trial and above a certain S/N is: 

\begin{equation}
    \label{eq:fakedetectionsingle}
    n_{\rm false}(> {\rm S/N} ) \simeq 2 N_s P(> {\rm S/N} ) \ , 
\end{equation}
where 
\begin{equation}
    \label{eq:cumulativeprob}
    P(>{\rm S/N}) = \int_{ {\rm S/N} }^{+\infty} \frac{1}{\sqrt{2 \pi}} e^{-\frac{x^2}{2}}  dx
            = \frac{1}{2} \left[1 - {\rm erf} \left( \frac{ {\rm S/N}}{\sqrt{2}} \right) \right] \ ,            
\end{equation}
and ${\rm erf}(x)$ is the error function. When processing a survey of $N_{\rm point}$ pointings of $N_{\rm beam}$ beams each and searching for bursts via a sub-banded search processing a total number of sub-bands $N_{\rm sub}$ the total number $N_{\rm false}(> {\rm S/N})$  of false detections above a certain S/N:

\begin{equation}
    \label{eq:fakedetections} 
     N_{\rm false}(> {\rm S/N}) \simeq 2 N_{\rm beam} N_{\rm point} N_s P(>{\rm S/N}) \Pi \left( N_{\rm DM} \right) \ , 
\end{equation}
where 
\begin{equation}
    \label{eq:htrupi}
    \Pi (N_{\rm DM}) = 
    \sum_{\rm all \ sub-bands}
    n_{\rm DM} \left(\nu_{\rm top}, \nu_{\rm bot}\right) 
\end{equation}  
is the total number of DM trials searched for each sub-band\footnote{The trial DM array is usually computed by compromising sensitivity loss/computing time. In this respect the sensitivity loss depends on the broadening effect of the burst width due to the choice of a wrong DM trial, which is, in turn, due to the dispersion delay between the two considered frequencies $\nu_{\rm bot}$ and $\nu_{\rm top}$, frequency dependent. This makes the size of the DM step dependent on the frequency window $ \left(\nu_{\rm top}, \nu_{\rm bot}\right)$. For a detailed discussion on this see, e.g., \cite{handbookpulsar}. } with top and bottom frequency $\nu_{\rm top}$ and $\nu_{\rm bot}$ respectively. 

Employing Eq.\,\ref{eq:fakedetections} and considering our survey specifics and the sub-banded search strategy, we expect less than 1 candidate to be detected by random chance due to noise for sub-banded ${\rm S/N} > 8.2$ (see Sect.\,\ref{sec:appendixb}). However, it is advisable to set a slightly higher S/N threshold than the computed value to account for the inherent uncertainty in S/N estimation, which has an error of 1 (see again Sect.\,\ref{sec:appendixb} for further details). This adjustment helps ensure that candidates close to the noise floor are not mistakenly considered as detections.

Additionally, the presence of RFI could significantly impact the effectiveness of the S/N threshold, as the unpredictable nature of RFI can lead to an increased number of false detections. Accurately quantifying how RFI might affect our results requires a comprehensive understanding of the specific RFI environment, which is beyond the scope of this work.

Another important factor to consider is the performance of the classifier. As indicated by the authors, {\sf FETCH} achieves a recall rate\footnote{Recall in binary classification represents the ratio of true events correctly identified by the classifier (True Positives, TP) to the sum of true positives and false events mistakenly identified as true events (False Positives, FP). Mathematically, Recall = TP / (TP+FN).} exceeding 99\% for S/N values greater than 10. Therefore, aligning the S/N threshold with this optimal range not only mitigates the effects of noise and RFI but also ensures that the classifier performs effectively. 



\begin{figure*}
    \centering
    \includegraphics[width = 0.95 \linewidth]{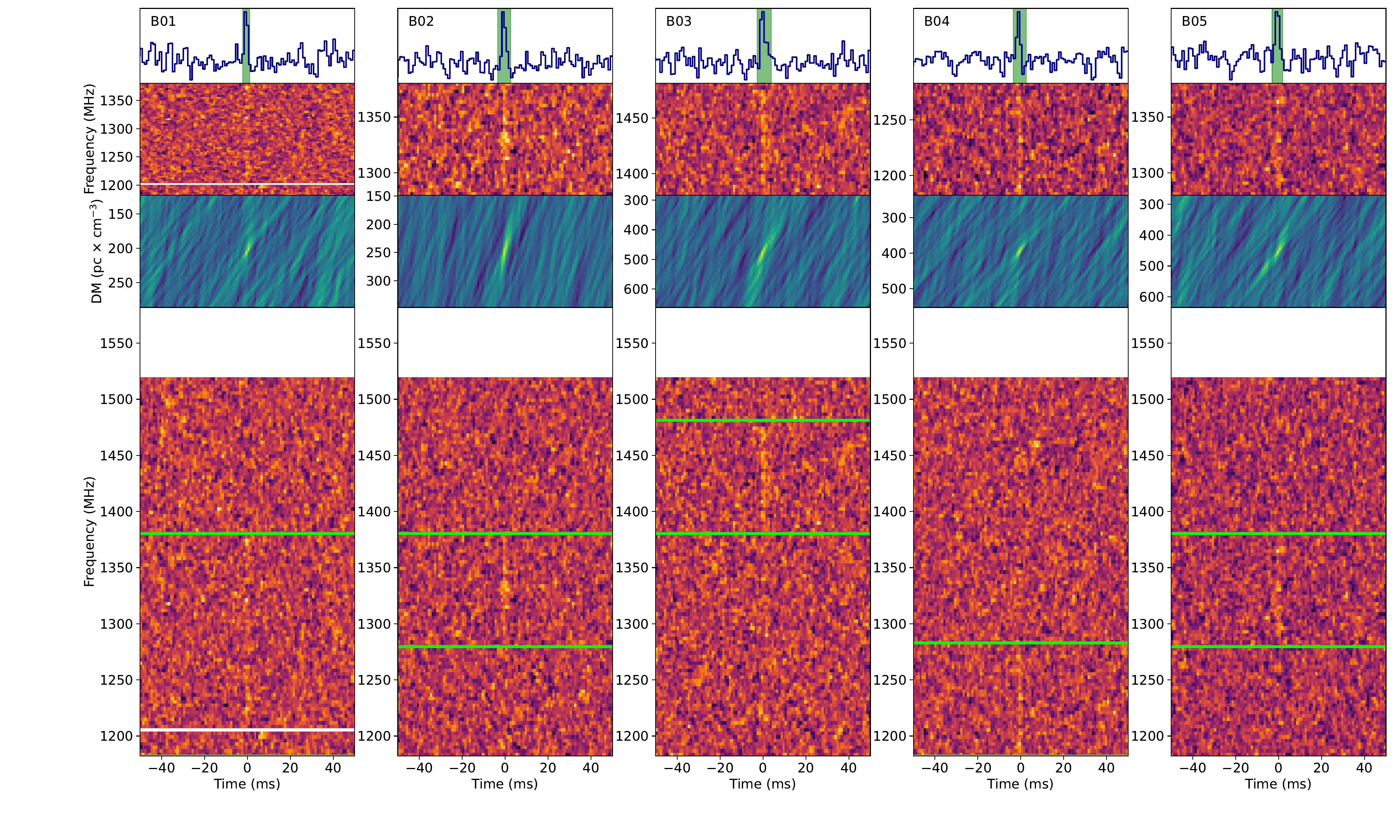}
    \caption[]{Narrow-band bursts detected in the HTRU HiLAT sub-band search. Here, we display the five highest sub-band S/N (see Sec,\ref{sec:appendixa} for the remaining candidates). For each burst, the bottom panel shows the dedispersed waterfall plot of the data, with two green lines delimiting the sub-band area in which the burst has been detected. The mid-lower panel shows the DM-time (butterfly) plot of the burst, and the mid-upper panel is the waterfall plot of the sub-banded dedispersed data. Lastly, the top panel shows the frequency-averaged time series along the sub-band. The blank rows in all the waterfall plots represent excised channels due to RFI.}
    \label{fig:frbsubs1}
\end{figure*}

Given this, we conservatively positively vetted candidates with ${\rm S/N} \geq 10$ (18 out of 51 of our candidates) as probable detections. The remaining candidates were rejected. Table\,\ref{tab:htruburstproperties} shows the properties of the detected bursts (in addition to the already discovered ones), named as BXX, and Fig.\,\ref{fig:frbsubs1} depicts the waterfall plots of the five highest sub-band S/N bursts (see Sect.\ref{sec:appendixc} for the remaining ones). The DM of these  candidates spans a range from a minimum of approximately $200$\,\pcc\ to a maximum of about $1650$\,\pcc, which is 2 to 30 times larger than the DM predictions for the Galactic contribution in their respective beam pointings \citep[obtained by using the software package {\sc pygedm}, ][]{pygedm}.
With these additional 18 detections, along with the 10 previous discoveries, the sub-banded search exhibited a detection gain of $180$\,\%. Despite being notably high for the $400$\,MHz band of the PMB, this value still aligns well with the predicted detection gain discussed in Sect.\,\ref{subsec:gain} for the PMB, falling within the $1 \sigma$ range error, further enhancing the significance of the candidates.

\subsection{Parameter distributions}

\begin{figure*}
    \centering
    \includegraphics[width = \linewidth]{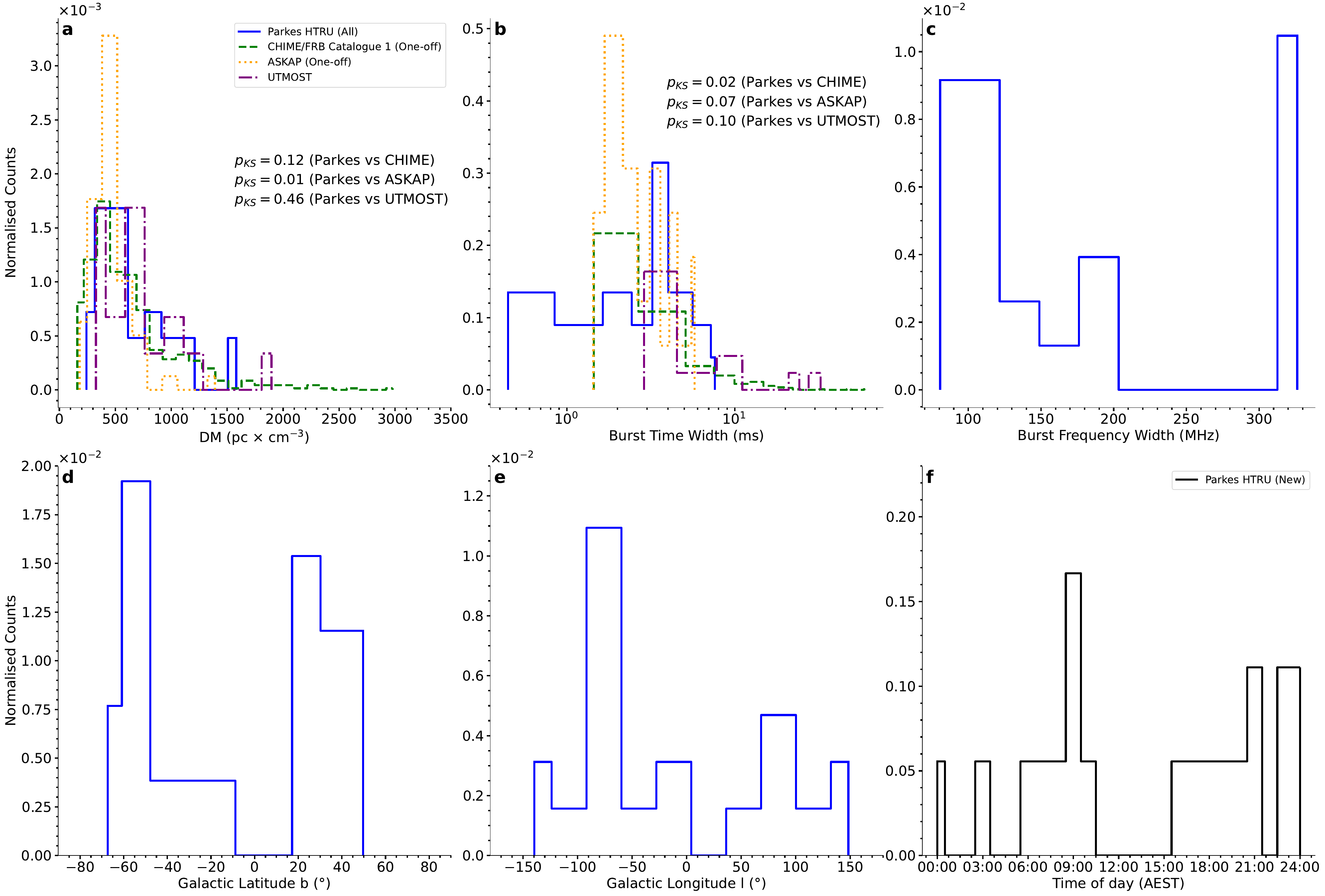}
    \caption[Distributions of the sub-banded detected bursts]{Distributions of DM (a), burst time width (b), burst frequency width (c), Galactic Latitude $b$ (d), and Galactic Longitude $l$ (e). The Parkes sample (blue) consists of our 18 detections along with the 10 already discovered FRBs in this survey \citep{htrufrb1,htrufrb2,htrufrb3}. For subplots (a, b), we also display, for comparison, the distribution obtained by considering the sample of the one-off FRBs published by CHIME \citep[green,][]{chime_cat1_21}, the distribution of the sample of ASKAP one-off FRBs \citep[orange,][]{shannon18, bannister19, macquart19, prochaska_2019_frblowdensity, bhandari_2023_oneoffdwarf, ryder_2023_frbz1}, and the sample of FRBs detected by UTMOST \citep[violet,][]{caleb_2017_utmost,farah_2019_utmost,gupta_2019a_atel,gupta_2019b_atel,gupta_2019c_atel,gupta_2020a_atel,gupta_2020b_atel,mandik_2021_atel,mandik_2022_atel}. For each pair of histograms in each subplot, we report the Kolmogorov-Smirnov probability $p_{KS}$ of being drawn from the same distribution. Panel (f) shows the distribution of the newly detected bursts arrival times in AEST local time.}
    \label{fig:htruhist}
\end{figure*}

\begin{figure}
    \centering
    \includegraphics[width = \linewidth]{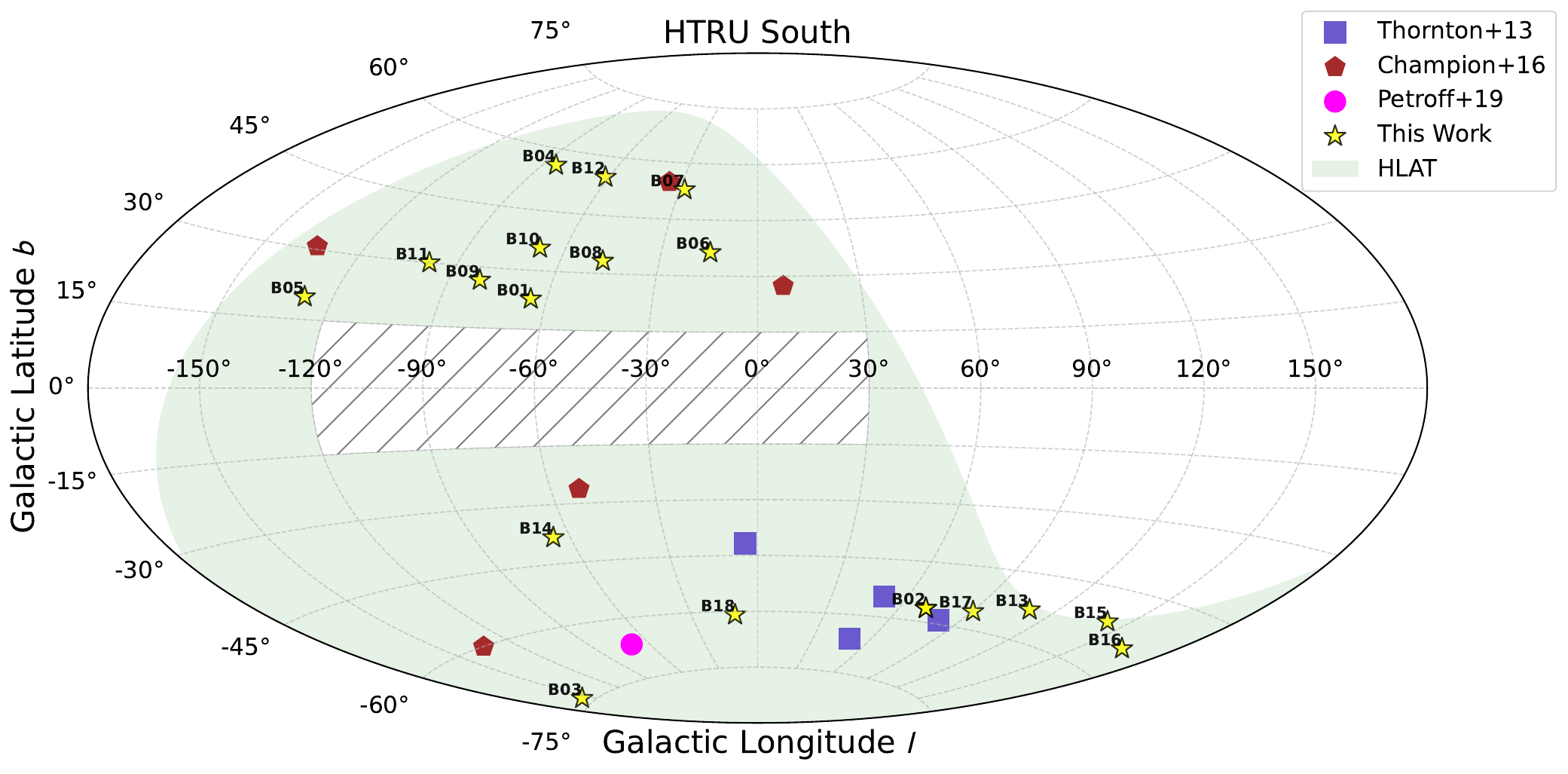}
    \caption[Sky Distribution of the FRBs discovered in HTRU]{Sky distribution (Aitoff projection) of the bursts detected in this work via the sub-banded search of HTRU HiLAT. For comparison, the bursts from \citet{htrufrb1,htrufrb2,htrufrb3} have also been plotted.}
    \label{fig:htruskyfrbs}
\end{figure}

Figure \ref{fig:htruhist} shows the parameter distribution of the HTRU bursts, which comprises the 18 new detections and the published bursts. As a comparison, for the DM and burst time width distribution, we also show the sample of the one-off FRBs published in the first CHIME/FRB catalogue \citep{chime_cat1_21}, the sample of one-offs published by ASKAP \citep{shannon18, bannister19, macquart19, prochaska_2019_frblowdensity, bhandari_2023_oneoffdwarf, ryder_2023_frbz1} and the sample of FRBs discovered by UTMOST \citep{caleb_2017_utmost,farah_2019_utmost,gupta_2019a_atel,gupta_2019b_atel,gupta_2019c_atel,gupta_2020a_atel,gupta_2020b_atel,mandik_2021_atel,mandik_2022_atel}. The data were retrieved from the on-line database FRBSTATS \citep{frbstats}. Each pair of histograms is accompanied by the Kolmogorov-Smirnov probability $p_{KS}$ that the two be drawn from the same distribution.

Regarding the DM distribution, from Fig.\,\ref{fig:htruhist}, we see that the Parkes HTRU distribution tends to peak at $\sim 500$ \pcc, with shape consistent with a log-normal distribution as similarly obtained by CHIME/FRB, ASKAP and UTMOST. This also applies for the time width distribution, with a peak around widths of $2-4$\,ms. In the case of the frequency width distributions we do not make a comparison with neither of the other samples as we followed, due to the sub-banded search, a different search strategy. The last two histograms of Fig \ref{fig:htruhist}, as well as Fig.\,\ref{fig:htruskyfrbs}, show the sky distribution of the beam pointings of the detected bursts.  The distributions are relatively uniform. We notice that there are less detections in the range $-15^{\circ} < b< 15^{\circ}$ consistently with how HiLAT was designed, as discussed in Sect.\,\ref{subsec:surveyspec}. 

In Fig.\,\ref{fig:htruhist} (panel f) we display the distribution of the time of arrivals, of our 18 detections, as local time in Australian Eastern Standard Time (AEST). This figure is done in order to check if our new sample could be possibly ascribed to the class of peryton signals \citep{peryton1,peryton2,peryton3}, an FRB-like signal which turned out to be RFI \citep{petroff_2015_perytons} from a microwave oven. Perytons are characterised as strong bursts (usually detected in multiple not adjacent beams of the PMB) frequency-sweeped closely to a $\nu^{-2}$ relation, giving them an ``apparent'' DM usually clustered around $\sim 400$\,\pcc or slightly less commonly around $\sim 200$\,\pcc. The ultimate test to establish their terrestrial nature was in fact to display their occurrence within the time of day noticing that they cluster around lunch and dinner time \citep[see,][]{petroff_2015_perytons}. From Fig.\,\ref{fig:htruhist} (panel f) we see that the distribution is fairly uniform, which in addition with our multi-beam exclusion criteria discussed in Sect.\,\ref{subsec:htrupipeline}, support the idea that it is unlikely that our candidates are weak perytons.

Lastly using the sample of 28 FRBs, following \cite{htrufrb1} and \cite{htrufrb2} we estimate an updated all-sky rate for HiLAT:
\begin{equation}
    \label{eq:htrurate} 
    R(>F_{\nu}  \sim 0.6 \ {\rm Jy \ ms}) = 
    4 \pi 
    \left( \frac{N_{\rm FRBs}}{0.0197} \right) \ {\rm sky^{-1} \ day^{-1}} \ \ , 
\end{equation}
which returns for $N_{\rm FRBs} = 28$ a rate of $1.8^{+0.4}_{-0.3} \times 10^4$\,sky$^{-1}$\,day$^{-1}$ \citep[with uncerteanties at 1$\sigma$\,c.l. assuming Poissonian statistics, ][]{gehrels_poissoncl}. The obtained value, scaled with respect to the Parkes completeness fluence of $\sim 2$\,Jy\,ms \citep{petroff_2015_complete} and assuming an Euclidean fluence slope of $-3/2$, is consistent with the Parkes all-sky rate reported by \cite{superb2}, being $1.7^{+1.5}_{-1.1} \times 10^3$\,sky$^{-1}$\,day$^{-1}$.

\section{Summary and conclusions}
\label{sec:conclusions}


This work provides a comprehensive analysis of sub-banded searches for fast radio bursts (FRBs), which involves detecting bursts within a specific portion of the full observational bandwidth spectrogram. 
Monte Carlo simulations show that a sub-banded search, aimed to detect bursts down to spectral extensions of $100$\,MHz, can yield a detection gain of $67_{-42}^{+133}$\,\% for the $400$\,MHz band Parkes 21-cm multi-beam (PMB) receiver system and of $1443_{-126}^{+143}$\,\% for the $3328$\,MHz band Parkes ultra-wideband low (UWL) receiver.

As a proof of concept we applied a sub-banded search to the high latitude portion of the HTRU South survey whose data were recorded with the PMB. Our results include the identification of eighteen new FRBs. This sample is almost twice the previously reported set of discoveries \citep{htrufrb1,htrufrb2,htrufrb3}, underscoring the value of sub-banded searches even for receivers with fairly limited bandwidths, such as the one of the PMB.

As radio telescope technology continues to progress, many of the latest receiver systems have the capability to process considerably wider observational bandwidths. Notable examples additional to the Parkes UWL are the ultra-wideband receiver of the Robert C. Byrd Green Bank Telescope \citep[GBT,][]{gbt_uwb} and the Ultra Broadband (UBB) receiver under commissioning at Effelsberg. This ongoing trend highlights the need for conducting burst searches within specific frequency ranges.

As these receiver systems incorporate larger bandwidths, alternative search strategies could come into play, such as the direct convolution of the spectro-temporal data matrix with a two-dimensional top hat function template rather than collapsing the bandwidth (or a portion of it in a sub-banded search) and searching for excesses in the time domain. However, it is worth noting that such an approach could present computational challenges, particularly with regard to achieving real-time detections, which could be required in modern facilities, due to the impracticality of storing all the data products they generate. In fact, considering a data matrix with $N_c$ frequency channels and $N_s$ time bins, a two-dimensional matched filter approach could have a time complexity of $\mathcal{O} (N_{\rm box}^t N_{\rm box}^{\nu} N_c N_s)$ when performing $N_{\rm box}^t , N_{\rm box}^{\nu}$ box-car trials in time and frequency, respectively. In contrast, a sub-banded search would require $\mathcal{O} ( N_{\rm sub} N_{\rm box}^t N_s)$ operations, where $N_{\rm sub}$ is the total number of sub-bands processed. In most scenarios, the sub-banded search approach would demand fewer computations and can be executed efficiently in parallel, further increasing the computational speed.

\begin{acknowledgements}
The authors thanks the anonymous referee for the useful comments which significantly improved the quality of this work. Part of the research activities described in this paper were carried out with the contribution of the NextGenerationEU funds within the National Recovery and Resilience Plan (PNRR), Mission 4 - Education and Research, Component 2 - From Research to Business (M4C2), Investment Line 3.1 - Strengthening and creation of Research Infrastructures, Project IR0000026 – Next Generation Croce del Nord. The Parkes (Murriyang) radio telescope is part of the Australia Telescope, which is funded by the Commonwealth Government for operation as a National Facility managed by CSIRO. The reprocessing of the HTRU-S HiLat survey made an extensive use of the GPU accelerators on OzStar supercomputer at Swinburne University of Technology. The OzSTAR program receives funding in part from the Astronomy National Collaborative Research Infrastructure Strategy (NCRIS) allocation provided by the Australian Government. M. T. expresses his immense gratitude to R. Humble for assistance with the OzSTAR cluster.
\end{acknowledgements}

\bibliographystyle{aa}
\bibliography{biblio}

\begin{appendix}

\section{Signal-to-noise ratio enhancement}
\label{subsec:htrusnr}

Let us consider a time series $f_n$ (with $n = 0,1,\dots,N -1$) of duration $T$. The series is then Fourier-transformed\footnote{We adopt the following convention for the Fourier transform: given a generic $N$-entries vector $x_n$, its Fourier transform corresponds to $y_n = \sum_{n = 0}^{N-1} \omega(N)^{\alpha}_{\ n} x^n$, where the projectors are $\omega(N)^{\alpha}_{\ n} = \exp \left( - i 2 \pi\frac{\alpha n}{N} \right)$. The power spectral density of $x_n$ is defined as $|y_n|^2$.} and modulus squared into a $N_c$ frequency channels and $N_s$ time bins spectrogram matrix  $F^{\alpha}_{\ l}$ (with $\alpha = 0,1,\dots,N_{c}-1 ; l = 0,1,\dots,N_s -1$); within the frequency range $\nu_c \pm {\rm BW}/2$, where $\nu_c$ is the observational central frequency and BW the observational bandwidth. We assume the spectrogram contains a candidate FRB and proceed to incoherently dedisperse the spectrogram at the correct DM value of the FRB candidate.  The matrix $F^{\alpha}_{\ l}$ can be decomposed into the sum of two matrices: 
\begin{equation}
    \label{eq:snrgain1} 
    F^{\alpha}_{\ l} = B^{\alpha}_{\ l} + N^{\alpha}_{\ l} \ , 
\end{equation}
where $B^{\alpha}_{\ l}$ contains only the FRB candidate signal and $N^{\alpha}_{\ l}$ is the noise matrix. 
Furthermore, we assume that the initial noise in the time series $f_n$ follows a Gaussian distribution with a zero mean and variance $\sigma^2$, i.e., $\in \mathcal{N}(0,\sigma^2)$. In the matrix $N^{\alpha}_{\ l}$, there will be $N_s$ Fourier power spectral densities of $N_c$ points, each following a gamma distribution $\Gamma \left(\kappa, \theta \right)$, with $\kappa = 1$ and $\theta = N_c \sigma^2$, which is therefore an exponential distribution, or equivalently, a chi-squared distribution with two degrees of freedom. This distribution possesses a mean and variance of $N_c \sigma^2$ \citep[see, e.g.,][\textsection\,9, for a proof]{blackmantukey_psd}. Further processing could alter the nature of the noise. For instance, decimating the spectrogram matrix, due to the central limit theorem, will render the noise Gaussian-distributed.

To describe the FRB signal in the spectro-temporal domain, contained in the matrix $B^{\alpha}_{\ l}$, it is more convenient to approximate $B^{\alpha}_{\ l}$ as a continuous real function $B(\nu,t)$, considering that $N_c >> 1$ and $N_s >> 1$. A reasonable first approximation model for a dedispersed FRB is a two-dimensional Gaussian function of the kind: 
\begin{equation}
    \label{eq:snrgain2} 
    B^{\alpha}_{\ l} \simeq B \left( \nu, t   \right) = 
    A e^{- \frac{ (\nu - \nu_0)^{2}  }{2 \sigma_\nu^2  }    } 
    e^{- \frac{ (t - t_0)^{2}  }{2 \sigma_t^2  }    } \ ,
\end{equation}
where $A$ represents the amplitude controlling the signal intensity, $\nu_0,\sigma_\nu$ are the central frequency of emission and standard deviation in frequency of the burst and $t_0, \sigma_t$ are the time of arrival and standard deviation in the time of occurrence of the burst. 

An important parameter for determining the significance of a burst is the S/N of the frequency-averaged profile, given by
\begin{equation}
\label{eq:snrgain3}
p_l = \frac{1}{N_c} \sum_{\alpha = 0}^{N_c -1} F^{\alpha}_{\ l} \ .
\end{equation}
For calculating the S/N of the profile $p_l$, we adopt the definition from \cite{handbookpulsar}:
\begin{equation}
\label{eq:snrgain4}
    {\rm S/N} = \frac{I}{\varepsilon_{\rm off} \sqrt { W_{\rm eq}}} \ ,
\end{equation}
where 
\begin{equation}
    \label{eq:intprofile}
    I = \sum_{l = 0}^{N_s} \left( p_l - \mu_{\rm off} \right)\ .
\end{equation}
Here, $\mu_{\rm off}$ and $\varepsilon_{\rm off}$ represent the mean and standard deviation of the off-pulse profile $p_l$, and $W_{\rm eq}$ is given by
\begin{equation}
\label{eq:snrgain5}
W_{{\rm eq}} = \frac{1}{ p_{z_{t_0}}}
\sum_{{\rm on \ pulse}} \left(p_l - \mu_{\rm off}\right) \ ,
\end{equation}
where $z_{t_0} = [t_0 / d t]$ corresponds to the discrete time bin of $t_0$. Our objective is to assess the potential S/N gain achievable by considering a narrow window $W_c$ (or $W_{\nu} = W_c d \nu $ in physical units, with $d \nu$ being the frequency resolution) that closely matches the observed frequency width of the FRB.

We now address the definition of physical widths for an FRB in time ($\Delta t$) and frequency ($\Delta \nu$) domains. Due to their complex structure \citep[see e.g.,][]{hessels19}, defining burst width is somewhat arbitrary. A common practice, assuming a Gaussian-like profile, employs the full width at half maximum (FWHM) in time and frequency. 
In sub-banding data for burst search, our aim is for window $W_\nu$ to closely approximate $\Delta \nu$.

From now on, we identify the quantities computed considering sub-band of the data using an asterisk $\ast$.
We want to evaluate how much the $\rm S/N^*$ improves in comparison to the full-band S/N. To make this computation, we will evaluate step-by-step the various elements in Eq.\,\ref{eq:snrgain4}. 

Let us start with the average profile $p_l$. Substituting Eq.\,\ref{eq:snrgain1} in Eq.\,\ref{eq:snrgain3}:
\begin{equation}
    \label{eq:snrgain7} 
    p_l = \frac{1}{N_c} \sum_{\alpha = 0}^{N_c -1} 
    \left (B^{\alpha}_{\ l} + N^{\alpha}_{\ l} \right) \ . 
\end{equation}
Separating the noise and signal, within the noise contribution we are summing $N_c$ random variables distributed as $\Gamma \left(1, N_c \sigma^2 \right)$. By exploiting properties of gamma-distributed random numbers we can write:
\begin{equation}
\begin{split}
    \label{eq:snrgain8} 
    \frac{1}{N_c} \sum_{\alpha = 0}^{N_c -1}  N^{\alpha}_{\ l}   
    & \sim  \frac{1}{N_c} \sum_{\alpha = 0}^{N_c -1} \Gamma \left( 1, N_c \sigma^2 \right) \\
    & \sim \frac{1}{N_c}  \Gamma \left( N_c, N_c \sigma^2 \right)  \\
    & \sim  \Gamma \left( N_c,  \sigma^2 \right) \sim \mathcal{N} \left(N_c \sigma^2, N_c \sigma^4 \right)  \ .
\end{split}    
\end{equation}

To obtain the result in Eq.\,\ref{eq:snrgain8} we exploited the following properties of the gamma distribution:  $ \sum_i \Gamma \left( \kappa_i, \theta \right)  \sim \Gamma \left( \sum_i \kappa_i, \theta \right)$ and $\Gamma \left( \kappa, \theta \right) \sim \mathcal{N} \left(\kappa \theta, \kappa \theta^2 \right) $ for $\kappa \rightarrow \infty $.

Making an abuse of notation, we evaluate now the burst term treating it as continuous function: 
\begin{equation}
\begin{split}
    \label{eq:snrgain9} 
    \frac{1}{N_c} \sum_{\alpha = 0}^{N_c -1} B^{\alpha}_{\ l} & \simeq 
    \frac{1}{\rm BW} \int_{\nu_c - {\rm BW}/2}^{\nu_c + {\rm BW} / 2} B \left(\nu,t \right) d \nu \\
    & \simeq
    \frac{1}{\rm BW} \int_{-\infty}^{+ \infty} B \left(\nu,t \right) d \nu 
    \\
    & = 
    \sqrt{2 \pi} A \frac{\sigma_\nu}{\rm BW} 
    e^{- \frac{(t-t_0)^2}{2 \sigma_t^2}} \ ,
\end{split}    
\end{equation}
where we extended the integral to the infinity since we assume that the burst frequency width is smaller than the observational bandwidth. The average profile in the full-band, $p_l$, is hence a Gaussian function in time, modelled as Eq.\,\ref{eq:snrgain9} with each time bin corrupted by a Gaussian noise $\in \mathcal{N} \left( N_c \sigma^2,  N_c \sigma^4 \right)$. %

In a similar fashion we can now compute the average profile considering a sub-band:
\begin{equation}
    \label{eq:snrgain10}
    p^*_l = \frac{1}{W_c} \sum_{\alpha = z_{\nu_0} - [W_c/2]}^{z_{\nu_0} + [W_c/2]} 
    F^{\alpha}_{\ l}
    = \frac{1}{W_c} \sum_{\alpha = z_{\nu_0} - [W_c/2]}^{z_{\nu_0} + [W_c/2]} 
    \left(
    B^{\alpha}_{\ l}  + N^{\alpha}_{\ l} \right)  \ ,
\end{equation}
where $z_{\nu_0} = [\nu_c / d \nu]$ is the corresponding discrete frequency channel of $\nu_0$. Handling the burst and the noise separately we can conclude analogously to what we did in Eq.\,\ref{eq:snrgain8} that the sub-banded profile noise will follow the statistics: 
\begin{equation}
    \label{eq:snrgain11} 
     \frac{1}{W_c} \sum_{\alpha = z_{\nu_0} - [W_c/2]}^{z_{\nu_0} + [W_c/2]} 
    N^{\alpha}_{\ l} \sim 
    \mathcal{N} 
    \left(N_c \sigma^2,  W_c \left(\frac{N_c}{W_c} \right)^2 \sigma^4 \right) \ ,
\end{equation}
and the sub-banded FRB profile can be obtained via the integral:
\begin{eqnarray}
\begin{split}
    \label{eq:snrgain12}
    \frac{1}{W_c} \sum_{\alpha = z_{\nu_0} - [W_c/2]}^{z_{\nu_0} + [W_c/2]} 
    B^{\alpha}_{\ l} & \simeq 
    \frac{1}{W_\nu}
    \int_{\nu_0 - W_\nu / 2}^{\nu_0 + W_\nu / 2} B \left(\nu,t \right) d \nu \\
    & \simeq  \sqrt{2 \pi} A \frac{\sigma_\nu}{W_\nu} 
    e^{- \frac{(t-t_0)^2}{2 \sigma_t^2}} 
\end{split}    
\end{eqnarray}
Similarly to the full-band profile,  we obtain a Gaussian function in time corrupted by a Gaussian noise.

As  the next step we show that the equivalent width is the same in both cases:
\begin{equation}
    \label{eq:weq} 
    W_{{\rm eq}} = \frac{1}{ p(z_{t_0})}
    \sum_{{\rm on \ pulse}} \left(p_l - \mu_{\rm off}\right) \simeq 
     \frac{1}{ p(z_{t_0})}
      p(z_{t_0}) \Delta t = \Delta t \ , 
\end{equation}
hence when considering the ratio between $\rm S/N^*$ and $\rm S/N$ that can be simplified. The standard deviations off-pulse of the profile are, thanks to Eqs.\,\ref{eq:snrgain8} and \ref{eq:snrgain11}:
\begin{equation}
    \label{eq:rmsoffpulse}
    \begin{split}
        \varepsilon_{\rm off}   & =  \sqrt{N_c} \sigma^2 \\
        \varepsilon^*_{\rm off} & =  \sqrt{W_c} \left(\frac{N_c}{W_c} \right) \sigma^2 \ .
    \end{split}
\end{equation}
Lastly, we need to evaluate the sum of the profile $I$. Again, assuming the profile as continuous function, we can approximate the sum as an integral: 
\begin{equation}
\begin{split}
    \label{eq:snrgain13} 
    I   \simeq \int_0^{T} p(t) dt 
     \simeq \int_{-\infty}^{+ \infty} p(t) dt  
    & = 
    \sqrt{2 \pi} A \frac{\sigma_{\nu}}{\rm BW} 
    \int_{-\infty}^{+ \infty} e^{- \frac{(t-t_0)^2}{2 \sigma_t^2}} dt \\
    & = 
    2 \pi A \frac{\sigma_\nu}{\rm BW} \sigma_t \ ,
\end{split}    
\end{equation}
in an analogue way, for the sub-banded profile:
\begin{equation}
    \label{eq:snrgain14}
    I^*  \simeq \int_0^{T} p^*(t) dt \simeq \int_{-\infty}^{+ \infty} p^*(t) dt = 
    2 \pi A 
    \frac{\sigma_\nu}{W_\nu} \sigma_t \ .
\end{equation}
Considering Eq.\,\ref{eq:snrgain4} for both profiles and taking the ratio:
\begin{equation}
\begin{split}
    \label{eq:snrgain} 
    \frac{\rm S/N^*}{\rm S/N} & = \frac{I^*}{I} \frac{\varepsilon_{\rm off}}{\varepsilon^*_{\rm off}}  
     \simeq \frac{N_c}{W_c} \frac{\varepsilon_{\rm off}}{\varepsilon^*_{\rm off}} \\ 
    & \simeq \sqrt{\frac{N_c}{W_c}} = \sqrt{\frac{\rm BW}{W_\nu}} \ . 
\end{split}    
\end{equation}

In conclusion, by selecting a sub-band whose width resembles the spectral extension of an FRB, a gain in S/N of the order of $\sqrt{ {\rm BW} / W_{\nu}}$ can be achieved. For instance, if a burst is confined to only the upper or lower half of the bandwidth and exhibits a S/N$\sim 7$, its S/N$^*$ can be approximately $\sqrt{2}$ times larger in its respective half-band. Consequently, the significance of the burst increases, yielding ${\rm S/N}^* \sim 10$.

Equation \ref{eq:snrgain} can be directly derived by the standard radiometer equation \citep{handbookpulsar}. The isotropic energy of the burst in the full-band is
\begin{equation}
    \label{eq:radiometer1} 
    E = \frac{4 \pi D^2_{\rm L}}{(1 + z)} S_{\nu} {\rm BW} \Delta t \ , 
\end{equation}
where $D_{\rm L}$, $z$ and $S_{\nu}$ are, respectively, the luminosity distance, the red-shift and the peak flux density of the burst. This energy will be the same in the sub-band $W_{\nu}$ where the burst resides,
\begin{equation}
    \label{eq:radiometer2} 
    E^* = \frac{4 \pi D^2_{\rm L}}{(1 + z)} S^*_{\nu} W_{\nu} \Delta t \ .
\end{equation}
The radiometer equation allows to translate the burst's S/N into physical unit of flux density:
\begin{eqnarray}
    \label{eq:radiometer3} 
    S_{\nu} &=& {\rm S/N} \frac{(T_{\rm sky} + T_{\rm sys})}{G \sqrt{n_p \Delta t {\rm BW}}} \\ 
    \label{eq:radiometer4} 
    S^*_{\nu} &=& {\rm S/N}^* \frac{(T_{\rm sky} + T_{\rm sys})}{G \sqrt{n_p \Delta t W_{\nu}}}
\end{eqnarray}
where $T_{\rm sky}$ is the sky temperature, $T_{\rm sys}$ is the system temperature, $G$ is the telescope gain and $n_p$ the number of polarisations. Replacing Eqs.\,\ref{eq:radiometer3},\ref{eq:radiometer4} in the respective sides we obtain Eq.\,\ref{eq:snrgain}.

It is noteworthy that if the noise matrix followed a Gaussian distribution, where each pixel is represented by a Gaussian random number with a  mean $\mu_N$ and variance $\sigma_N^2$, Eq.\,\ref{eq:snrgain} would still remain valid. This assertion can be verified by following the analogous steps as demonstrated in Eqs.\,\ref{eq:snrgain8} and \ref{eq:snrgain10}. Specifically, if $N^{\alpha}_{\ l} \in \mathcal{N}(\mu_N, \sigma_N^2)$, then $\varepsilon_{\rm off} = \sigma_N / \sqrt{N_c}$ and $\varepsilon^*_{\rm off} = \sigma_N / \sqrt{W_c}$, thus obtaining again the same relationship between S/N and S/N$^*$.

To perform this calculation, we employed a number of significant approximations, ranging from assumptions about the noise's nature to the Gaussian-bell shape of the burst and the complete absence of other spurious signals. Consequently, Eq.\,\ref{eq:snrgain} will rarely hold in real-world scenarios. However, when considering the observational bandwidth of the receiver and the burst spectral occupancy of interest, it offers a reasonable metric for evaluating the potential gain in S/N achievable through a sub-banded search.

From Eq.\,\ref{eq:snrgain}, it becomes evident that the burst from FRB\,20190711A, detected with the UWL receiver as reported by \cite{kumarsub}, would have eluded detection without a sub-banded search. The reported S/N in its specific sub-band is $\sim 12$, which, given its spectral extension relative to the full bandwidth, would correspond to a full-band S/N of approximately $\sim 1.7$. This value falls considerably below the detection thresholds employed by search pipelines, due to the lack of statistical significance and the remarkably large number false positives to be inspected.

When the burst fully spans the observational bandwidth, where $\sigma_{\nu} \sim {\rm BW}$, we can make the assumption that $\sigma_{\nu} \gg 1$. Under this assumption, we can  Taylor expand at the first order Eq.\,\ref{eq:snrgain2} with respect to $ (\nu - \nu_0) / \sigma_{\nu}$, yielding:

\begin{equation}
    \label{eq:snrloss} 
    B(\nu , t) \simeq A e^{-\frac{(t-t_0)^2}{2 \sigma_t^2}} \ ,
\end{equation}

This implies that in each frequency channel, we have a Gaussian function in time, as modeled by Eq.\,\ref{eq:snrloss}. Consequently, the expressions for $I$ and $I^*$ in Eqs.\,\ref{eq:snrgain13},\ref{eq:snrgain14} become equivalent since the profiles $p(t)$ and $p^*(t)$ are the same. However, the off-pulse standard deviation of the noise is still described by Eq.\,\ref{eq:rmsoffpulse}. Therefore, we arrive at:

\begin{equation}
    \label{eq:snrlossf}
   \frac{{\rm S/N}^*}{\rm S/N} 
   = \frac{I^*}{I} \frac{\varepsilon_{\rm off}}{\varepsilon^*_{\rm off}} 
   = \frac{\varepsilon_{\rm off}}{\varepsilon^*_{\rm off}} = \sqrt{\frac{W_\nu}{\rm BW}}  \ .
\end{equation}
That is, when the burst fully encompasses the observational bandwidth, we have a S/N loss rather than a gain.

\section{Signal-to-noise ratio threshold}
\label{sec:appendixb} 

We derive the minimum S/N required to make noise false alarms negligible. Suppose the noise is Gaussian with a mean of zero and a standard deviation of one. If we search for significant excesses over $\rm N_{trial}$ trials, the number of noise false alarms will be simply ${\rm N_{trial}} P(>{\rm S/N})$, where $P(>{\rm S/N})$ is given by Eq.\,\ref{eq:cumulativeprob}. To ensure a negligible number of noise false alarms, we impose the condition:
\begin{equation}
\label{eq:noisefalse1} 
{\rm N_{trial}} P(>{\rm S/N}) < 1 \ .
\end{equation}
Substituting Eq.\,\ref{eq:cumulativeprob} and inverting the error function yields:
\begin{equation}
\label{eq:snrthreshold}
{\rm S/N}_{\rm thresh} > \sqrt{2} \ {\rm erf}^{-1}(z) \ ,
\end{equation}
where
\begin{equation}
\label{eq
}
z = \frac{ \rm N_{trial} - 2}{ \rm N_{trial}} \ .
\end{equation}
Equation\,\ref{eq:snrthreshold} can be directly computed using standard Python packages such as SciPy \citep{scipy}. Additionally, for a large number of trials where $z \sim 1$, Eq.\,\ref{eq:snrthreshold} can be asymptotically expanded obtaining a good approximation \citep{errorfunctioninverse}:
\begin{equation}
\label{eq:noisefalse2} 
{\rm S/N}_{\rm thresh} > \sqrt{2 \eta - \ln \eta } \ ,
\end{equation}
where
\begin{equation}
\label{eq:snrthresholdapproximate} 
\eta = - \ln \left[ \sqrt{\pi} \left( 1 - z \right) \right] \ .
\end{equation}
Equation\,\ref{eq:snrthreshold} provides a reasonable value for the S/N of a candidate to be unlikely noise, thereby defining the "noise floor." However, after computing this S/N threshold, it is prudent to raise it slightly. This adjustment accounts for two important factors.

First, the S/N itself has an inherent error of 1, meaning that even under ideal conditions (e.g., Gaussian noise and perfect signal matching), some events might be missed purely due to this uncertainty. For instance, a strict threshold at 10$\sigma$ could result in missing 50 \% of events with a "real" S/N (that is the one obtained by the radiometer equation) of 10, 16 \% of those with an S/N of 11$\sigma$, and so on.

Second, the presence of non-Gaussian noise and spurious RFI signals in real data further complicates detection. By setting the threshold slightly above the theoretical noise floor, one can reduce the likelihood of false detections caused by RFI, improving the overall reliability of the detected signals.

While Equation\,\ref{eq:snrthreshold} offers a good baseline, raising the threshold helps to compensate for the inherent uncertainties in S/N estimation and the challenges posed by real-world observational data.

\section{Properties of the bursts and remaining waterfall plots}
\label{sec:appendixc}

\begin{sidewaystable*} 
\caption[]{Properties of the 10 rediscovered HTRU FRBs adn the 18 bursts discovered.}
\label{tab:htruburstproperties}
    \centering
    \begin{tabular}{lcccccccccccl}
    \hline
    \hline
    ID & TOA$|^{\nu = \nu_{\rm top}}_{\rm topo}$ & R.A. & Dec. & DM & DM$_{\rm Gal}$ & $\Delta t$ & Sub-band & $\nu_0$ &  $\Delta {\nu}$ & S/N$^*$ & $F_{\nu}$ &  \\ 
     & [MJD] & [hh:mm:ss] & [$^{\circ}:^{\prime}:^{\prime \prime}$] & [\pcc] &  [\pcc] & [ms] & [MHz] & [MHz]  & [MHz] &  &  [Jy ms] & \\ 
    \hline
    \hline
FRB\,20110222A &  55612.080428 & 22:34:00 & 12:24:00 & 944.3(8) & 34.40  & < 5.6 & (1182-1582) & 1382 & > 340 & 49  & $>8.0$ & [1] \\
FRB\,20110626A  & 55739.898119 & 21:03:00 & -44:44:00 & 727.0(3) & 50.0  & <1.4 & (1182-1582) & 1382 & > 340 &  11  & $>0.7$ & [1]  \\
FRB\,20110703A  & 55745.791442 & 23:30:00 & -02:52:00 & 1103.6(7) & 31.6  & < 4.3 &  (1182-1582) & 1382 & > 340 & 16  & $>1.8$ & [1]  \\
FRB\,20120127A  & 55953.341224 & 23:15:00 & -18:25:00 & 553.3(3) & 32.3  & 1.1 & (1182-1582) & $1202 \pm 14$ & $116 \pm 18$ &  11  & $>0.6$ & [1]  \\
FRB\,20090625A  & 55007.912417 & 03:07:47 & -29:55:36 & 899.6(1) & 32.0  & < 1.9 & (1182-1582) & 1382 & > 340  & 29  & $>2.2$ & [2]  \\ 
FRB\,20121002A  & 56202.548131 & 18:14:47 & -85:11:53 & 1628.18(2) & 74.0 & 0.3 & (1182-1582) & 1382 & > 340 &  16  & $>2.3$ & [2]  \\ 
FRB\,20130626A  & 56469.622223 & 16:27:06 & -07:27:48 & 952.4(1) & 67.0  & 0.12 & (1182-1582) & 1382 & > 340 &  20  & $>1.5$ & [2]  \\
FRB\,20130628A  & 56471.165278 & 09:03:02 & 03:26:16 & 469.88(1) & 53.0  & < 0.05 & (1182-1582) & 1382 & > 340 &  29  & $>1.2$ & [2]  \\ 
FRB\,20130729A  & 56502.376304 & 13:41:21 & -05:59:43 & 861(2) & 31.0 & < 4.0 & (1182-1582) & $1219 \pm 2$ & $47 \pm 4$ &  14  & $>3.5$ & [2]  \\
FRB\,20110214A  & 55606.301505 & 01:21:17 & -49:47:11 & 168.8(5) & 31.10  & < 1.9 & (1182-1582) & $1202 \pm 3$ & $49 \pm 5$ &  13  & $>1.1$ & [3] \\
B01 &  56488.386616 & 12:15:30 & -39:35:19 & 204.4(5) & 84.50  & < 3.3  & (1182-1382) & $1288 \pm 20$ & $197 \pm 90$ &  12  & $>0.7$ & [4] \\
B02 &  55672.962508 & 23:08:08 & -04:34:28 & 248.0(5) & 33.20  & < 6.1  & (1282-1382) & $1340 \pm 9 $ & $82 \pm 25$ &  11  & $>0.9$ & [4] \\
B03 &  55561.338024 & 02:01:16 & -23:38:22 & 472.7(5) & 29.50  & < 6.6  & (1382-1482) & $1434 \pm 5 $ & $67 \pm 15$ &  11  & $>1.0$ & [4] \\
B04 &  56662.790473 & 11:51:05 & -03:22:38 & 394.8(3) & 34.20  & < 6.1 & (1182-1282) & $1233 \pm 15$ & $73 \pm 30 $ &  11  & $>0.9$ & [4] \\
B05 &  56405.326536 & 08:37:35 & -06:49:43 & 452.7(7) & 72.10  & < 5.0 & (1282-1382) & $1295 \pm 9$  & $163 \pm 20$ &  11  & $>0.7$ & [4]\\
B06 & 56653.889492 & 15:08:09 & -15:04:36 & 527.4(6) & 47.50  & < 4.7  & (1282-1382) &  $1308 \pm 10$ & $ 68 \pm 20 $ & 11 &  $>0.8$  & [4] \\
B07 & 56656.906695 & 13:59:28 &  -06:04:42 & 458.5(6) & 170.40  & < 3.9 & (1382-1582) & $1458 \pm 5$  & $76 \pm 10$ &  11 &  $>0.7$ & [4] \\
B08$^*$ & 56675.962208 & 13:29:20 & -28:52:35 & 1078.6(9) & 57.60  & < 4.9  & (1382-1482) & 1432  & 100 &  11 & $>0.7$ & [4] \\
B09 & 56657.668195 & 11:10:48 & -31:19:44 & 535.5(5)  & 73.80 & < 4.3  & (1182-1282) & $1224 \pm 9$  & $75 \pm 20$ &  11 &  $>0.6$ & [4] \\
B10$^*$  & 56632.742044 & 12:12:06 & -26:18:36 & 396.5(7) & 33.50  & < 3.8  & (1282-1382) & 1332  & 100 &  11 & $>0.7$ & [4] \\
B11 & 56585.800067 & 10:21:11 & -20:33:52 & 1655.1(3) & 56.40  & < 7.0  & (1382-1482) & $1453 \pm 9$  & $73 \pm 20$ & 11 & $>0.9$ & [4] \\
B12$^*$ & 56469.443399 & 12:41:18 & -08:01:30 & 410.7(6) & 34.70  & < 3.7  & (1282-1382) & 1332 & 100 &  11 & $>0.6$ & [4] \\
B13 & 55981.277722 & 00:20:17 &	 08:55:11 & 766.3(3) & 32.90  & < 3.6  & (1382-1482) & $1424 \pm 25 $  & $125 \pm 100 $ &  10 & $>0.6$ & [4] \\
B14 & 56124.013309 & 02:50:35 & -76:50:34 & 681.5(3) & 47.70  & < 8.0  & (1282-1382) & $1337 \pm 10$ &  $94 \pm 40$ &  10 & $>0.9$ & [4] \\
B15 & 56027.130313 & 01:39:46 & 09:23:19 & 534.8(2) & 35.80  & < 2.8 & (1282-1382)  &   $1360 \pm 30$ & $102 \pm 55$ &  10 & $>0.6$ & [4] \\
B16$^*$ & 55592.376355 & 02:20:52 & 00:46:01 & 546.0(6) & 35.50  & < 3.9  & (1282-1382) & 1332  & 100 &  10 & $>0.7$ & [4] \\
B17 & 56296.408940 & 23:40:57 & 02:33:27 & 371.3(8) & 33.00  & < 2.1  & (1382-1582)  & $1436 \pm 20$  & $131 \pm 40$ &  10 & $>0.6$ & [4] \\
B18 & 56140.860458 & 22:53:07 &-44:00:37 & 397.0(1) & 33.90  & < 2.8  & (1382-1482) &  $1443 \pm 20$ & $118 \pm 75 $ &  10 & $>0.6$ & [4] \\ 
    \hline
    \end{tabular}
\tablefoot{
Each burst is assigned an ID in the first column. The second column contains the time of arrival (TOA) of the candidates, computed with respect to the topocentric reference frame and at a frequency $\nu_{\rm top} = 1582$\,MHz. Columns three and four contain the Right Ascension (R.A.) and Declination (Dec.) beam pointings where the burst occurred. The fifth column represents the DM of the bursts, while the sixth column contains the DM prediction of the Galactic contribution based on the NE2001 \citep{ne2001} or YMW16 \citep{ymw17} models, whichever is the highest for that pointing. $\Delta t$ is the FWHM time width of the bursts, computed via a Gaussian fit, corrected for intra-channel smearing and sampling time. As the scattering contribution is not resolved for most bursts, these widths are reported as upper limits. The eighth column reports the discovery sub-band. $\nu_0$ and $\Delta \nu$ in the ninth and tenth columns are the frequency centroid and FWHM width, respectively, obtained via a Gaussian fit. For the FRBs that cover the full observational band, the frequency centroid has been assigned to the central frequency of the PMB, and the observational bandwidth is reported as the lower limit to the effective bandwidth of the burst. Bursts for which the Gaussian frequency fit did not converge have the detection sub-band central frequency and width reported instead. S/N$^*$ represents the signal-to-noise ratio computed with respect to the discovery band for each burst. $F_{\nu}$ is the fluence computed via the radiometer equation. Note that these values should be considered as lower limits, as we assumed the telescope gain at the central beam of the PMB in the radiometer equation. References: [1] \cite{htrufrb1}, [2] \cite{htrufrb2}, [3] \cite{htrufrb3}, [4] This work.
}
\end{sidewaystable*}

\begin{figure*}
    \centering
    \includegraphics[width = 0.95 \linewidth]{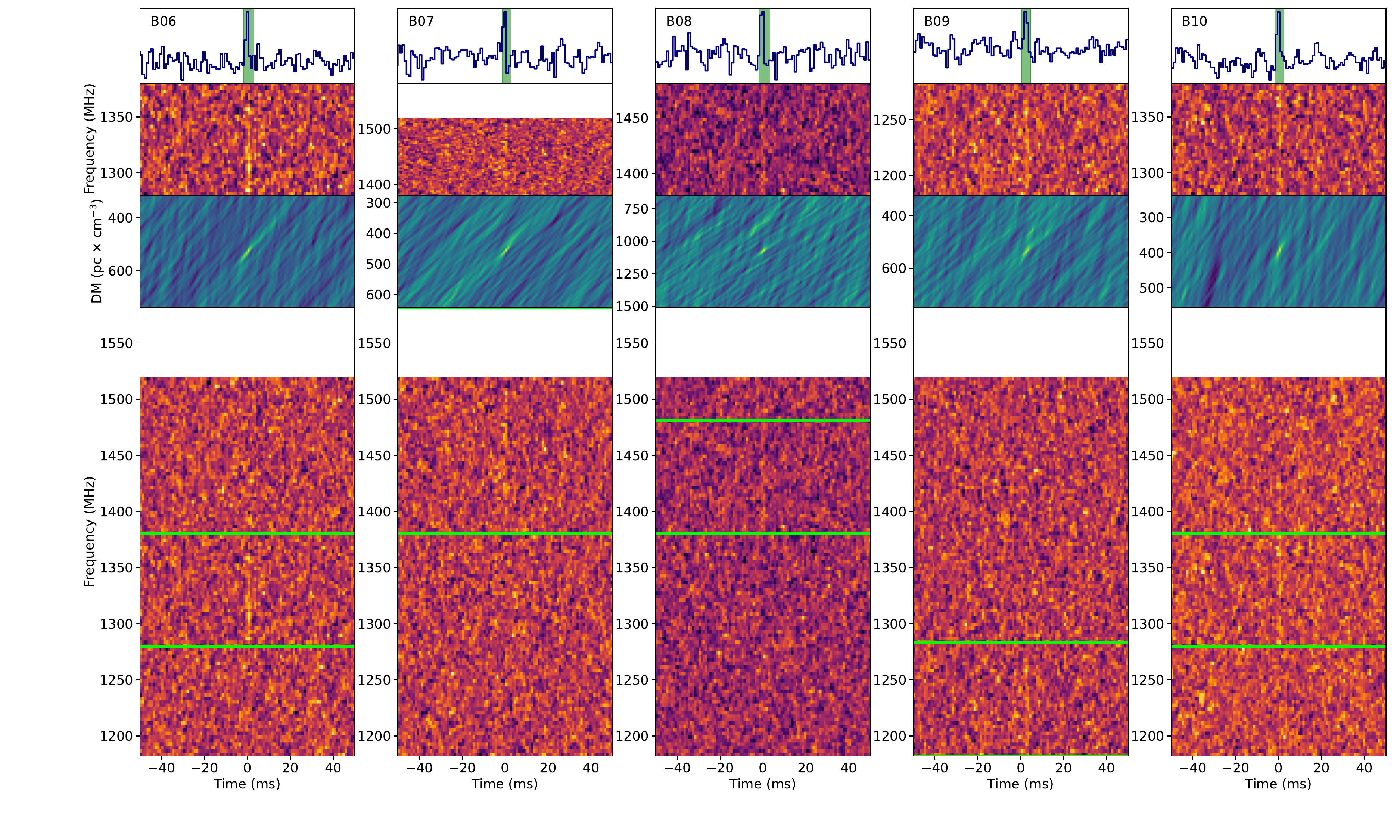}
    \caption[]{The same as Fig.\,\ref{fig:frbsubs1} for B06-10.}
    \label{fig:frbsubs2}
\end{figure*}

\begin{figure*}
    \centering
    \includegraphics[width = 0.95 \linewidth]{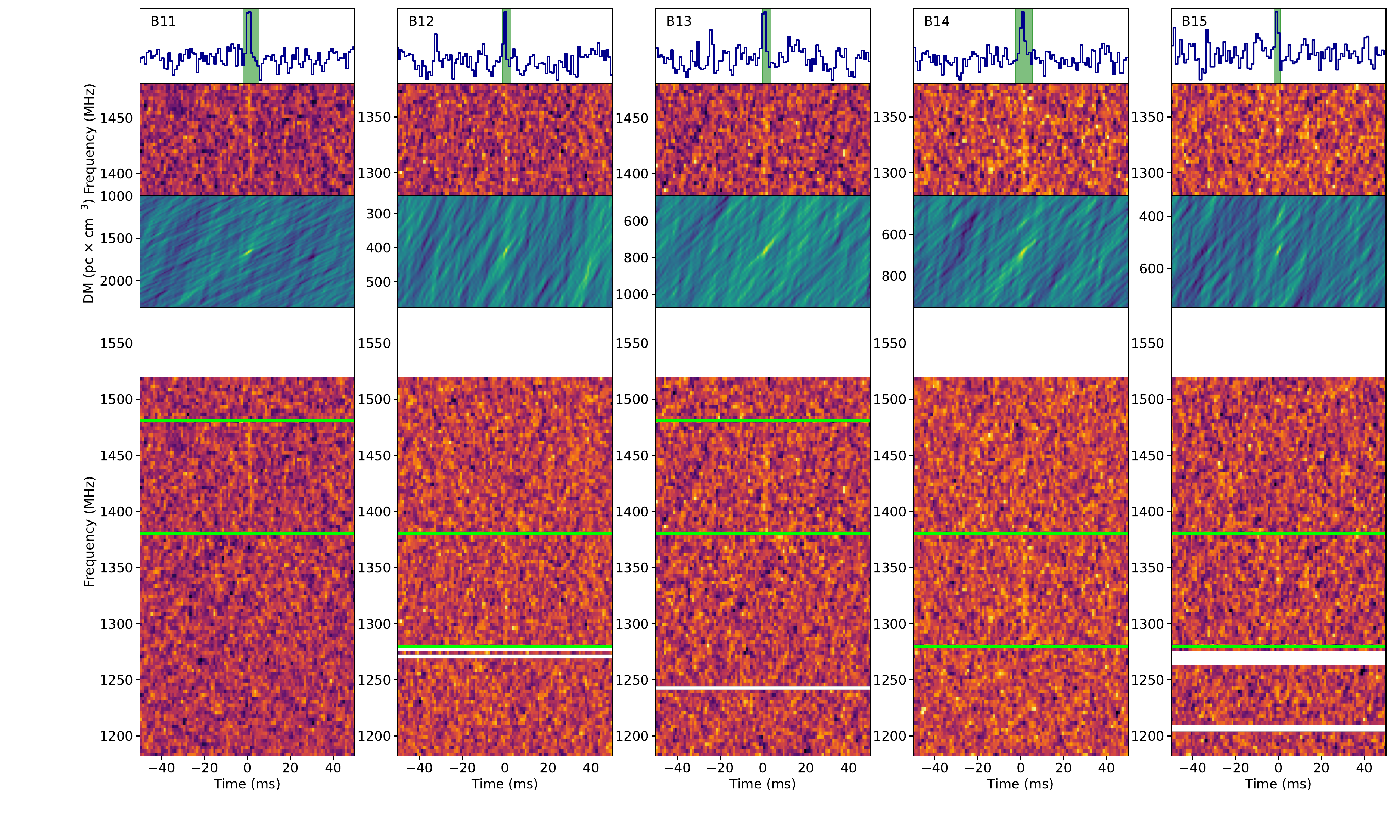}
    \caption[]{The same as Fig.\,\ref{fig:frbsubs1} for B11-15.}
    \label{fig:frbsubs3}
\end{figure*}

\begin{figure*}
    \centering
    \includegraphics[width = 0.95 \linewidth]{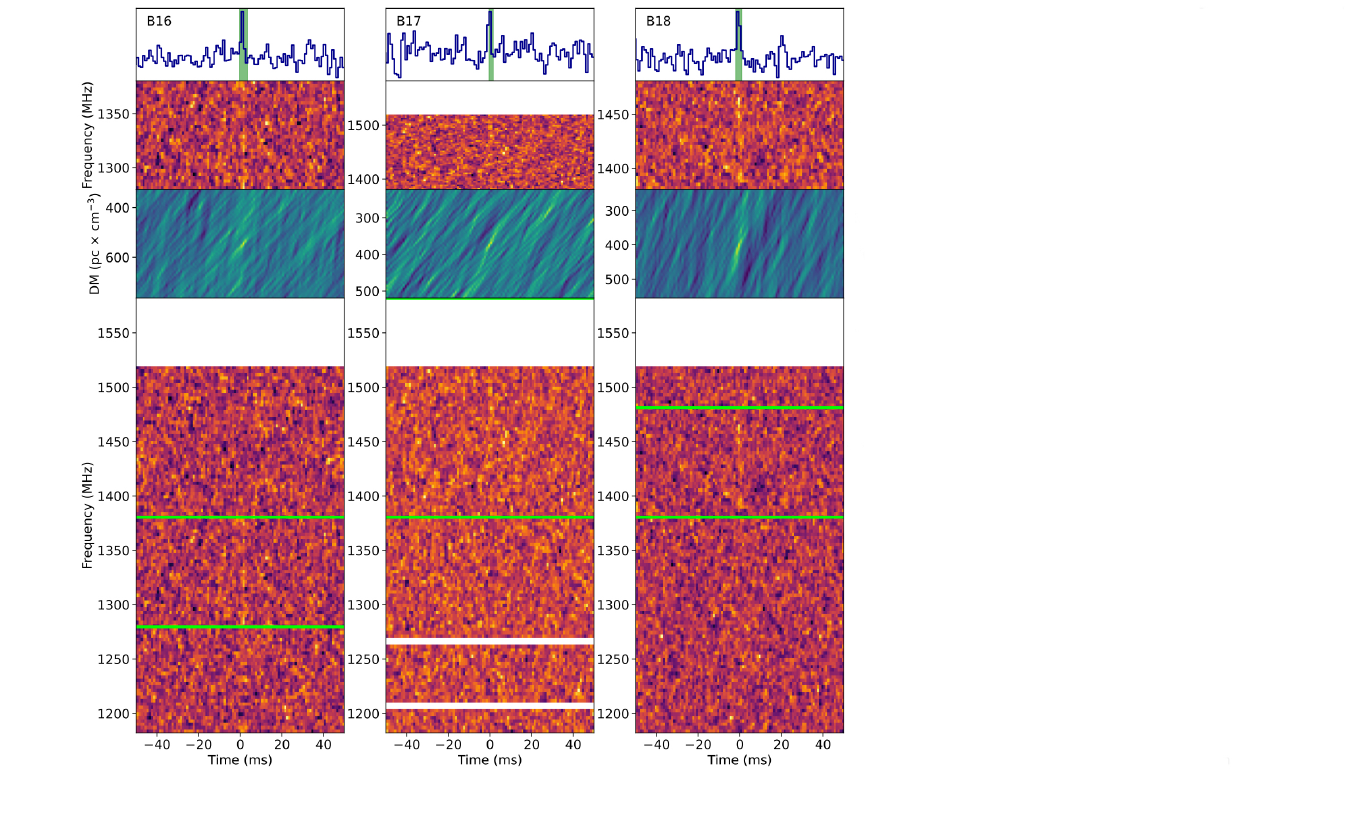}
    \caption[]{The same as Fig.\,\ref{fig:frbsubs1} for B16-18.}
    \label{fig:frbsubs4}
\end{figure*}








\end{appendix}

\end{document}